\documentclass[10pt,twocolumn]{IEEEtran}
\usepackage{epsfig,latexsym}
\usepackage{float}
 \usepackage{enumerate}
\usepackage{indentfirst}
\usepackage{amsmath}
\usepackage{amssymb}
\usepackage{times}
\usepackage{subfigure}
\usepackage{cite}
\usepackage{url}
\usepackage{color}
\usepackage{setspace}

\definecolor{green}{rgb}{0.796,0.948,0.816}
\def\Re{\textrm{Re}}

\begin{document}
\title{Beamforming Techniques for Non-Orthogonal Multiple Access in 5G Cellular Networks}
\author{Faezeh~Alavi,~\IEEEmembership{Student Member,~IEEE,}  Kanapathippillai~Cumanan,~\IEEEmembership{Member,~IEEE,} Zhiguo~Ding,~\IEEEmembership{Senior Member,~IEEE,} and Alister~G. Burr,~\IEEEmembership{Member,~IEEE,}
\thanks{F. Alavi, K. Cumanan and A. G. Burr are with the Department of Electronic Engineering, University of York, York, YO10 5DD, U.K.
(e-mail: \{sa1280, kanapathippillai.cumanan, alister.burr\}@york.ac.uk).
\mbox{Z. Ding} is with the School of Electrical and Electronic Engineering, The University of Manchester, Manchester, M13 9PL, U.K. (e-mail: zhiguo.ding@manchester.ac.uk).
}
}
\maketitle
\begin{abstract}
In this paper, we develop various beamforming techniques for downlink transmission for multiple-input single-output
(MISO) non-orthogonal multiple access (NOMA) systems.
First, a beamforming approach with perfect channel state information (CSI) is investigated
to provide the required quality of service (QoS) for all users.
Taylor series approximation and semidefinite relaxation (SDR) techniques are employed to reformulate the original non-convex
power minimization problem to a tractable one.
Further, a fairness-based beamforming approach is proposed through a max-min formulation to maintain fairness between users.
Next, we consider a robust scheme by incorporating channel uncertainties,
where the transmit power is minimized while satisfying the outage probability requirement at each user.
Through exploiting the SDR approach, the original non-convex problem is reformulated in a
linear matrix inequality (LMI) form to obtain the optimal solution.
Numerical results demonstrate that the robust scheme can achieve better performance compared
to the non-robust scheme in terms of the rate satisfaction ratio.
Further, simulation results confirm that NOMA consumes a
little over half transmit power needed by OMA for the same data rate requirements.
Hence, NOMA has the potential to significantly improve the system performance
in terms of transmit power consumption in future 5G networks and beyond.

\end{abstract}

\vspace{-0.3cm}
\section{Introduction}
The exponential growth of mobile data and
multimedia traffic imposes
high data rate requirements in the next generation wireless networks \cite{6692652,7676258,7891617,6730679}.
In handling this enormous amount of data traffic, multiple access techniques
play a crucial role
through efficiently accommodating multiple users \cite{6730679,7263349,6381677,6704653,7986959}. 
Recently, non-orthogonal multiple access (NOMA) has been envisioned as one of the key enabling techniques to address
these high data rate requirements and it is expected to significantly enhance
throughput as well as to support massive connectivity in 5G networks and beyond.
{Conventional wireless transmission employs orthogonal multiple access (OMA)
techniques in which orthogonal resources such as time, frequency and code are assigned to different users to remove inter-user interference.
Although this approach allows simple transceiver
implementations, it comes at the cost of spectral and energy efficiency. 
{NOMA outperforms conventional multiple access schemes such as
time division multiple access (TDMA), \cite{7272042,7971926},
orthogonal frequency division multiple access (OFDMA) \cite{7146204}, and zero-forcing (ZF) \cite{dingMIMO,7277111} by simultaneously sharing the available
communication resources (i.e., frequency and time) between all users via the power or code domain multiplexing which offers
a significant performance gain in terms of spectral efficiency \cite{6692652,7676258}.}

NOMA allocates more transmit power to the users with poor channel conditions
whereas the users with better channel conditions are served with less transmit power.
Then, successive interference cancellation (SIC) is applied at the receivers to efficiently
remove the interference caused by the weaker users.
The principle of superposition coding with SIC can be related to the concept of cognitive radio systems \cite{5665898,7332326}.
In particular,
NOMA allows controllable interference and allocates non-orthogonal resources
to increase system throughput
while introducing a reasonable additional complexity at the receiver \cite{7263349}.

In comparison with conventional user scheduling
which prefers to allocate more power to the users with better channel gains and increase the overall system
throughput but exacerbate unfairness,
NOMA enables a more flexible management of the achievable rate of the users and provides better fairness. In fact, it facilitates a balanced tradeoff between system throughput and user fairness \cite{whitepaper}. 
{In the literature, there are two types of NOMA scheme considered:
I) clustering NOMA \cite{6735800,7015589,7442902}, 
II) non-clustering NOMA \cite{7922522,8292467}.
In the clustering NOMA scheme, all the users in a cell are grouped into $N$ clusters with at least two users
in each cluster, for which a transmit beamforming vector is designed to support each cluster
through conventional multiuser beamforming designs. The users in each cluster are supported by a NOMA based
beamforming approach. However, in the non-clustering NOMA scheme, there is no clustering and each
user is supported by its own NOMA based beamforming vector.
Clustering is generally employed in a NOMA system with a large number of users to reduce the
computational complexity of the SIC. In addition, the performance of a beamforming design in cluster-based
NOMA system mainly depends on how the users are grouped into a number of clusters with different number of
users in each cluster. It is important to point out that the resulting combinatorial optimization problem
is in general NP-hard, and performing an exhaustive search for an optimal solution is computationally prohibitive.}

Recently, the NOMA scheme has received considerable
attention in research community due to its potential benefits in 5G and beyond networks.
In \cite{6868214}, the NOMA scheme was studied for downlink transmission {in} a cellular system
with randomly deployed users whereas the design of uplink NOMA schemes has been proposed in \cite{6933459}.
In \cite{dingPairing} a hybrid multiple access system
has been presented by combining NOMA with the conventional multiple access scheme,
where the impact of user pairing on the performance of NOMA systems is studied.
In \cite{7117391}, a novel cooperative NOMA scheme has been proposed, giving derivations of the
outage probability and diversity order.
{Joint power allocation and relay beamforming design is investigated in \cite{7833194} for a NOMA based amplify-and-forward relay network where the achievable rate of the destination with the best channel condition is maximized with rate requirements at other destinations and individual transmit power constraints.}

Multiple antenna techniques offer potential benefits in wireless communications through their additional spatial degrees of freedom \cite{7417680,7510972,7784829}, which
can be exploited to further enhance the performance of NOMA. 
{The authors in \cite{7277111} have investigated a beamforming design to maximize sum rate in a
multiple-input single-output (MISO) NOMA system using the CCP method which is mainly based on the Taylor series approximation
to transform the non-convex constraints into convex form.
Joint optimization with beamforming design and power
allocation for clustering MISO-NOMA systems is considered in \cite{7917241} where
an iterative algorithm is proposed based on semidefinite relaxation (SDR) to minimize power, showing that this algorithm requires less transmit power than for power allocation and beamforming considered
separately.}
In \cite{7833022}, a secure beamforming design is proposed for a MISO NOMA
system by grouping the users into clusters.

A general framework for multiple-input multiple-output (MIMO) NOMA systems has been studied in \cite{dingMIMO} by exploiting
signal alignment for both downlink and uplink transmission.
In \cite{7236924}, a design for precoding and detection has
been developed for MIMO-NOMA by deriving the outage
probabilities for two different power allocation schemes whereas the MIMO-NOMA network with limited feedback is considered in \cite{7434594}. The optimal and low
complexity power allocation schemes have been proposed in \cite{7095538} for a two-user MIMO-NOMA system.
In \cite{6692307}, a NOMA scheme has been proposed for downlink transmission of a MIMO system by employing intra-beam
superposition coding of multiple user signals and an intra-beam
SIC. In this scheme,
the number of transmitter beams is restricted to the number of
transmitter antennas, which is the same as in OMA in LTE-Advanced systems.
For the MIMO NOMA system, the secrecy rate maximization problem is
solved in \cite{7937906} where it is demonstrated that the NOMA scheme outperforms the conventional OMA scheme
in terms of achieved sum secrecy rate by efficiently utilizing available bandwidth.

In most existing NOMA schemes, it is assumed that perfect channel state information (CSI) is available at the transmitter,
however, in wireless transmissions, channel uncertainties are inevitable due to quantization and channel estimation errors,
limited training sequences and feedback delays. Particularly, due to ambiguities introduced in SIC through user decoding order
and superposition coding at the transmitter in NOMA, these uncertainties can greatly degrade the overall system performance.
Therefore, to cultivate the desirable benefits offered by NOMA, these channel uncertainties should be accounted for in
the design of resource allocation techniques.
To circumvent the inevitable channel uncertainties, robust design is a well-known approach in the literature, which can be
classified into two groups, the worst-case robust design \cite{4895658,6156468,7177089,7547398}, 
and the stochastic robust design \cite{6831610,6503905}. 
In the worst-case design, it is assumed that the CSI errors belong to some known bounded uncertainty sets and robust beamforming
design is proposed to tackle the worse error whereas in the stochastic approach, the channel errors are random with a certain
statistical distribution and constraints can be satisfied with certain outage probabilities. In fact, the bounded robust
optimization is generally conservative owing to its worst-case criterion while probabilistic SINR constrained beamforming
provides a soft SINR control. In the context of NOMA, a robust design with the norm-bounded channel uncertainties is studied
in \cite{7442902,7922522}. In \cite{7442902}, a clustering scheme was studied to maximize the worst-case achievable sum rate
with a total transmit power constraint whereas the non-clustering NOMA approach with a dedicated beamformer was developed for
a robust power minimization problem in \cite{7922522}.

In this paper, we consider a downlink MISO-NOMA system with a small number of users for which a number of beamforming techniques have been developed. The contributions of the proposed designs are summarized as follows:
\begin{itemize}
  \item Sum-power Minimization: Low energy consumption is one of the key requirements in future wireless networks.
  As such, we first consider a power minimization problem where each user should be satisfied with a predefined quality of service (QoS),
  which is measured in terms of minimum rate requirements. This scenario could arise in a network consisting of
  users with delay-intolerant real-time services (real-time users) \cite{otters}.
  These users should achieve their
  required QoS at all times, regardless of channel conditions. To solve this power minimization problem in a standard
  MISO NOMA system, we applied two different approaches: I) Taylor series approximation and II) SDR to design the
  beamforming. Furthermore, we evaluated the performance of these approaches in terms of transmit power consumption
  and computational complexity while comparing their performance with that of the conventional OMA schemes.
  \item Max-min fairness: For the previously considered power minimization approach, the transmitter requires a certain
  amount of transmit power to achieve the target rate at each user. However, the maximum available transmit power is
  generally limited at the transmitter and therefore the power minimization problem might turn out to be infeasible due
  to insufficient transmit power. In this case, the target rate should be decreased, and optimization should be repeatedly
  performed until the problem becomes feasible. To overcome this infeasibility issue, a max-min fairness based approach
  is considered 
  in which the minimum rate between all users is maximized while satisfying the transmit power constraint.
  This practical scenario could arise in a network consisting of users with delay-tolerant packet data services
  (non-real time users) \cite{1262126,5280202,6169204},
  where packet size could be varied according to the achievable rate value.
  In the context of NOMA systems, there are a number of works that consider the max-min fairness scheme in single antenna
  NOMA systems \cite{7069272,8254179}. However, this has not been studied for a multi-antenna NOMA system. In this work,
  we consider the max-min problem in a MISO NOMA system to maintain fairness between users. Unfortunately, this max-min problem
  is non-convex and the corresponding solution cannot be easily obtained. Hence, we first transform the problem into a convex
  one and utilize a bisection method to obtain the optimal solution for the original max-min fairness problem. In addition,
  simulation results have been provided to demonstrate the effectiveness of this max-min beamforming design.
  \item Robust design: In the context of NOMA, a robust design with norm-bounded channel uncertainties is studied in
  \cite{7442902,7922522}. However, in the worst-case design, extremely conservative approaches are considered which might
  result in higher transmit power consumption to meet the required QoS, whereas the probability-based design requires less
  transmit power to satisfy the outage probability constraints. Accordingly, we have considered an outage probabilistic
  based robust beamforming design by incorporating channel uncertainties, which has not been studied for the MISO NOMA system
  in the literature.
\end{itemize}

The rest of the paper is organized as follows. Section II describes the system model
whereas beamforming design through the power minimization problem  is presented in Section III with two different approaches.
The user fairness based max-min problem is formulated and solved in Section IV. The robust beamforming approach is proposed through
incorporating channel uncertainties in Section V. In Section VI, simulation results are provided to validate the effectiveness of
the proposed schemes. Finally, Section VII concludes this paper.

\subsection*{Notations}
Throughout this paper, we use lowercase boldface letters for vectors and uppercase boldface letters for matrices. {$(\cdot)^T$, $(\cdot)^H$ and ${\rm Tr(\cdot)}$ denote transpose, conjugate transpose and the trace of a matrix, respectively.
${\rm Pr(\cdot)}$ and $ \mathbb{E}\{\cdot\}$  stand for the probability operator and statistical
expectation for random variables, respectively.}
The symbols $\mathbb{C}^n$ and $\mathbb{R}^n_+$ are used for n-dimensional complex and nonnegative real
spaces, respectively. $ \mathbf{A} \succeq \mathbf{0}$ indicates that
$\mathbf{A}$ is a positive semidefinite matrix and
${\rm vec}(\mathbf{A})$ is the
vector obtained by stacking the columns of $\mathbf{A}$ on top of one
another.
$\Re(.)$ and $\Im(.)$  stand for the real and imaginary parts
of a complex number, respectively. $\mathbf{I}$ denotes the identity matrix with appropriate size and $\odot$ indicates Hadamard product.
 The Euclidean norm of a matrix is denoted by $\|\cdot\|_2$.
The notation $[\cdot]_{mn}$ represents the $mn^{th}$ element of a matrix.
$\mathcal{N}$ and $\mathcal{CN}$ stand for real and complex Gaussian random variable, respectively.

\section{System Model}
We consider a downlink transmission
for $N$ single antenna users, $U_k$, $k\in \{1,\ldots,N\}$. 
The base station (BS), equipped with $M$ antennas, exploits NOMA to simultaneously transmit signals to different users.
In particular, the BS transmits a superposition of the individual messages, i.e., $\mathbf{w}_k s_k$,
to all users, where $s_{k}$ and $\mathbf{w}_{k}$ are the
symbol intended for $U_{k}$ $\big(\mathbb{E}(|s_k|^2)=1\big)$ and the corresponding beamforming, respectively.
{Note that $\|\mathbf{w}_k\|_2^2$ represents the transmit power assigned to user $U_{k}$.}
The received signal at $U_{k}$, is given by
\begin{equation}\label{signal}
  y_k = \mathbf {h}_k^H \mathbf{w}_k s_k + \sum_{m\neq k}\mathbf{h}_k^H \mathbf {w}_m  s_m + n_k,
\end{equation}
where $\mathbf{h}_k \in \mathbb{C}^M$ denotes the complex channel vector between the BS and the $k^{th}$ user and $n_{k}$ represents zero-mean circularly symmetric additive white Gaussian noise with variance $\sigma^{2}$ at user $U_{k}$, (i.e., $n_k \sim \mathcal{CN }(0, \sigma^2)$).


{Assume that
users are ordered based on their channel quality i.e., $\|\mathbf{h}_{1}\|_2\leq\|\mathbf{h}_{2}\|_2\leq\ldots\leq\|\mathbf{h}_{N}\|_2$.
NOMA exploits the power domain to transmit multiple signals over the same frequency and time domain, and performs SIC at the receivers
to decode the corresponding signals \cite{5522468,6736749}.
Based on this ordering, each user, $U_k$, can detect and remove the first $k-1$ users' signals in a
successive manner whereas the message of the other users, i.e., from $U_{k+1}$ to $U_{N}$, is treated as noise. In other words, the $k^{th}$ user's signal should
be detected by $U_l$ for all $l \in \{k, k+1, \ldots , N\}$ \cite{6704653,7277111}.
{We should mention here that this ordering may not be optimal, and better rates may
be achievable for different decoding order of the users \cite{7277111}. However,
our work in this paper does not focus on the optimal decoding ordering problem, but in
the robust design of the beamforming vectors that minimize
the total transmit power of the system, for a given user ordering.}
Hence, the remaining signal at $U_l$ to detect
the $k^{th}$ user is represented as follows:}

\begin{align}\label{signal2}
  y^l_k = \mathbf {h}_l^H \mathbf{w}_k s_k + \sum_{m=k+1}^{N}\mathbf{h}_l^H \mathbf {w}_m  s_m + n_l,
\end{align}

Based on these conditions,
the achievable rate of the $k^{th}$ user can be obtained as follows:

\begin{align}\label{SINR}
R_k=\log_2 \big(1+\min_{l \in \{k, k+1, \ldots, N\}} ~ \text{SINR}_k^l\big),
\end{align}
where
\begin{equation}\label{SINR2}
 \text{SINR}^l_k= \dfrac{|\mathbf{h}_l^H \mathbf{w}_{k}|^2}{\sum_{m=k+1}^{N}|\mathbf{h}_l^H \mathbf{w}_{m}|^2+\sigma^2},
\end{equation}
denotes the SINR of the signal intended for the $k^{th}$ user at $U_l$.
Moreover, the following conditions should be satisfied in the NOMA scheme to guarantee the intended ordering of SIC in decoding the signals of the weaker users \cite{7277111}.
\begin{align}\label{nomacons} \nonumber
|\mathbf{h}^H_k \mathbf{w}_1|^2\geq\ldots\geq |\mathbf{h}^H_k \mathbf{w}_{k-1}|^2\geq|\mathbf{h}^H_k \mathbf{w}_k|^2\geq|\mathbf{h}^H_k \mathbf{w}_{k+1}|^2\geq \\
\qquad\qquad\qquad\qquad\qquad \ldots\geq|\mathbf{h}^H_k \mathbf{w}_N|^2, \quad \forall k.
\end{align}

In SIC based receivers, each user decodes its own message after decoding the messages of weaker users and
successfully removing their interference. In order to facilitate this SIC technique, the received power of the signals
to be decoded should be made greater than the received powers of the other users' signals. Hence, the above inequalities
are defined to implement SIC by increasing the power of the signals intended for the weaker users.
Through imposing these conditions, the users located far from the BS (cell-edge users) receive more signal power
than that of the users near to the BS.

\section{Power Minimization}
{In this section, we consider power minimization problem to satisfy throughput requirements at each user {where it is assumed that }the perfect CSI is available at all nodes.
This power minimization problem can be formulated into the following optimization framework:}
\begin{subequations}\label{mainproblem}
\begin{align}
  \min_{\mathbf{w}_k\in \mathbb{C}^{M\times 1}}  & \sum_{k=1}^{N} \| \mathbf{w}_k\|^2_2, \\\label{const1.1}
  s.t.  & \log_2 \big(1+\min_{l \in \{k, k+1, \ldots, N\}} ~ \text{SINR}_k^l\big)\geq R_k^{min}, \forall k.
\end{align}
\end{subequations}
{recalling that $\|\mathbf{w}_k\|^2_2$ represents the transmit power assigned to $U_k$} and the constraint in \eqref{const1.1} represents the minimum rate requirement $R_k^{min}$ at $U_k$.
{Since the SINRs required to successfully implement SIC are satisfied through
the minimum rate constraints in problem \eqref{mainproblem}, the constraints in \eqref{nomacons} become unnecessary in the design.}

This power optimization problem is non-convex and cannot be directly solved to realize the solution.
To tackle this issue, we exploit two different approaches to approximate the original problem and convert it into equivalent formulations.
Before presenting a detailed treatment of our approaches, we start with some transformations to simplify the constraints.
Since $\log(.)$ is a non-decreasing function, the constraint in \eqref{const1.1} can be represented as follows:
\begin{equation}\label{minconst}
\min_{l \in \{k, k+1, \ldots, N\}} ~ \text{SINR}_k^l \geq \gamma^{min}_k, \qquad \forall k,
\end{equation}
where
$\gamma^{min}_k=2^{R_k^{min}}-1$ is the minimum required SINR at $U_k$. {Without loss of generality,
the above constraint in \eqref{minconst} can be easily} rewritten as follows:
\begin{align}\nonumber
&\left\{\begin{array}{l l} \text{SINR}_k^k \geq \gamma^{min}_k, \\
\text{SINR}_k^{k+1} \geq \gamma^{min}_k, \\
\vdots \\
\text{SINR}_k^N \geq \gamma^{min}_k,
\end{array}
\right.
\end{align}

\begin{align}\nonumber
&\Leftrightarrow \left\{\begin{array}{l l} \gamma^{min}_k (\sum_{m=k+1}^{N}|\mathbf{h}_k^H \mathbf{w}_{m}|^2+\sigma^2)  \leq |\mathbf{h}_k^H \mathbf{w}_{k}|^2 ,  \\
\gamma^{min}_k (\sum_{m=k+1}^{N}|\mathbf{h}_{k+1}^H \mathbf{w}_{m}|^2+\sigma^2)  \leq |\mathbf{h}_{k+1}^H \mathbf{w}_{k}|^2, \\
\vdots \\
\gamma^{min}_k (\sum_{m=k+1}^{N}|\mathbf{h}_{N}^H \mathbf{w}_{m}|^2+\sigma^2)  \leq |\mathbf{h}_{N}^H \mathbf{w}_{k}|^2,  \end{array}
\right.
\end{align}

\begin{align}\nonumber
&\Leftrightarrow \gamma^{min}_k (\sum_{m=k+1}^{N}|\mathbf{h}_{l}^H \mathbf{w}_{m}|^2+\sigma^2)  \leq |\mathbf{h}_{l}^H \mathbf{w}_{k}|^2, \\ \label{const1} &\qquad\qquad \qquad\qquad\qquad \forall k,~ l=k,\ldots,N.
\end{align}

Finally, the equivalent formulation of the original power minimization problem \eqref{mainproblem} can be reformulated as
\begin{subequations}\label{mainproblem2}
\begin{align}
\min_{\mathbf{w}_k\in \mathbb{C}^{M\times 1}}  & \sum_{k=1}^{N} \| \mathbf{w}_k\|^2_2, \\\nonumber
s.t.\quad  &  \gamma^{min}_k (\sum_{m=k+1}^{N}|\mathbf{h}_{l}^H \mathbf{w}_{m}|^2+\sigma^2)  \leq |\mathbf{h}_{l}^H \mathbf{w}_{k}|^2, \\\label{const2.1}
&\qquad\qquad \forall k,~ l=k,\ldots,N.
\end{align}
\end{subequations}

\subsection{Non-Convex Constraint Approximation}
In this subsection, we provide convex approximations for non-convex constraints.
{To this end, first we consider the constraint \eqref{const2.1}
and equivalently transform it into a tractable form.
In this beamforming design, choosing arbitrary phase for $\mathbf{w}_{k}$ will not have any impact on the optimization and
will also provide the same solutions. Thus, any arbitrary phase can be selected for this beamformer. Furthermore, this enables us
to assume that $\mathbf{h}_l^H \mathbf{w}_{k}>0$, which makes the square root of $|\mathbf{h}_l^H \mathbf{w}_{k}|^2$
well-defined \cite{emilbook,8187586}. 
By reshuffling the constraint and taking the square root, we can
reformulate the non-convex constraints as a second-order cone (SOC) and linear constraints as follows:
\begin{align}\nonumber
& \gamma^{min}_k (\sum_{m=k+1}^{N}|\mathbf{h}_l^H \mathbf{w}_{m}|^2+\sigma^2)  \leq |\mathbf{h}_l^H \mathbf{w}_{k}|^2 \\ \label{SOC}
&\Leftrightarrow \left\{ \begin{array}{l l}
\sqrt{\gamma^{min}_k} \left\| \begin{array}{l l} |\mathbf{h}^H_l \mathbf{w}_{k+1}|  \\
\vdots\\
|\mathbf{h}^H_l \mathbf{w}_{N}|\\ \quad \sigma\\
\end{array}
\right\| \leq |\mathbf{h}_l^H \mathbf{w}_{k}|, \\
\\
\Im(\mathbf{h}_l^H \mathbf{w}_{k})=0
\end{array}
\right.
\end{align}
}

{However, it is impossible to have a phase rotation to simultaneously satisfy the following conditions:
\begin{equation}
  \Im(\mathbf{h}_k^H \mathbf{w}_{k})=\Im(\mathbf{h}_l^H \mathbf{w}_{k})=0, ~~~\forall l=k+1,\cdots,N.
\end{equation}
Therefore, we have applied this phase rotation only to satisfy $\Im(\mathbf{h}_l^H \mathbf{w}_{k})=0$, for  $l=k$, and exploited the Taylor series approximation \cite{6168880,6626661,7100916} for $l=k+1,\cdots,N$ to convexify the non-convex constraints in \eqref{const2.1} based on the following Lemma:}

\vspace{0.5cm}
\textit{Lemma 1:}
By using the first order Taylor series approximation of the function $f_l(\mathbf{w}_{k})$ around $\mathbf{w}^t_{k}$ in $t^{th}$ iteration, it holds that
 \begin{align}\nonumber
& |\mathbf{h}_l^H \mathbf{w}_{k}|^2 = \mathbf{w}_{k}^H \mathbf{h}_l \mathbf{h}_l^H \mathbf{w}_{k} \triangleq f_l(\mathbf{w}_{k}) \geq
 \\ \nonumber
 &{\mathbf{w}^t_{k}}^H \mathbf{h}_l \mathbf{h}_l^H \mathbf{w}^t_{k}
 + 2 \Re[{\mathbf{w}^t_{k}}^H \mathbf{h}_l{\mathbf{h}_l}^H(\mathbf{w}_{k}-\mathbf{w}^t_{k})]
\triangleq g_l(\mathbf{w}_{k},\mathbf{w}^t_{k}).
 \end{align}

This approximation is linear in terms of $\mathbf{w}_{k}$ and will be used instead of the original norm-squared function.
All inequality constraints in \eqref{const2.1} for $l=k+1,\cdots,N$ will be replaced by the following approximated convex constraints:
\begin{align}
& \gamma^{min}_k (\sum_{m=k+1}^{N}|\mathbf{h}_l^H \mathbf{w}_{k}|^2+\sigma^2)  \leq g_l(\mathbf{w}_{k},\mathbf{w}^t_{k}).
\end{align}
\begin{IEEEproof}
 Please refer to Appendix \ref{App.A}.
\end{IEEEproof}

Based on the SOC representation in \eqref{SOC} and the approximation in Lemma 1, the following optimization problem is formulated:
\begin{subequations}\label{taylor}
\begin{align}
 \min_{\mathbf{w}_k\in \mathbb{C}^{M\times 1}} & \sum_{k=1}^{N} \| \mathbf{w}_k\|^2_2, \\ \nonumber
 s.t.\quad & \left\{\begin{array}{l l}  \sqrt{\gamma^{min}_k} \left\|\begin{array}{l l} |\mathbf{h}^H_k \mathbf{w}_{k+1}|  \\
\vdots\\
|\mathbf{h}^H_k \mathbf{w}_{N}|\\ \quad \sigma\\
\end{array}
\right\| \leq |\mathbf{h}_k^H \mathbf{w}_{k}|, \\
\Im(\mathbf{h}_k^H \mathbf{w}_{k})=0,
\end{array}
\right. \\ \label{const3.1}
& \qquad\qquad\qquad\qquad\qquad\qquad\qquad\qquad \forall k,
\\ \nonumber
& \gamma^{min}_k (\sum_{m=k+1}^{N}|\mathbf{h}_l^H \mathbf{w}_{k}|^2+\sigma^2)  \leq g_l(\mathbf{w}_{k},\mathbf{w}^t_{k}),\\ \label{const3.3}
&\qquad\qquad\qquad\qquad \forall k, \:l=k+1,\ldots,N.
\end{align}
\end{subequations}

An iterative algorithm is developed to solve the power minimization problem based on the approximated problem in \eqref{taylor} which is
summarized in Table \ref{tayloralgorithm}.
This algorithm will be initialized with $\mathbf{w}^t_{k}$ and the corresponding approximated problem will be solved to obtain the beamforming vector, i.e., $\mathbf{w}^{t+1}_{k}$. In other words, the corresponding initial solution is updated iteratively and the algorithm will be terminated {once the required accuracy is achieved.}

\begin{table}[t] \caption{Taylor series approximation}\label{tayloralgorithm}
\centering
\begin{small}
\begin{tabular} {|l|}
\hline
\textbf{Algorithm 1.}  ~~Proposed Algorithm for solving problem \eqref{taylor}\\
\hline
1. \textbf{Initialization:}~ Set $t=0$ and randomly generate a set of\\
 \qquad \qquad \qquad \qquad\!\!\!\! feasible $\mathbf{w}^0_k ~\forall k$ for problem in \eqref{taylor}. \\
2. \textbf{repeat}\\
3. \qquad\quad Solve problem \eqref{taylor}\\
4. \qquad\quad Update $\{\mathbf{w}^{t+1}_k\}=\{\mathbf{w}^t_k\}$,\\
5. \qquad\quad $t \leftarrow t+1$ \\
6. \textbf{until}\quad$|\mathbf{w}_k^{t+1}-\mathbf{w}_k^{t}|\leq \varepsilon$ \\  
\hline
\end{tabular}
\end{small}
\end{table}

\subsection{Semidefinite relaxation approach}
Here, we provide another scheme to solve the original non-convex power minimization problem in \eqref{mainproblem2}.
By considering \mbox{$\mathbf{H}_k=\mathbf{h}_k \mathbf{h}_k^H$} and
$\mathbf{W}_k=\mathbf{w}_k \mathbf{w}_k^H$, a new matrix variable
$\mathbf{W}_k$ is introduced and the original power minimization problem
in \eqref{mainproblem2} can be reformulated as:
\begin{subequations}\label{rankone}
\begin{align}
 \min_{\mathbf{W}_k\in \mathbb{C}^{M\times M}}  & \sum_{k=1}^{N} {\rm Tr}(\mathbf{W}_k) \\ \nonumber
s.t.\quad  & \gamma^{min}_k (\sum_{m=k+1}^{N}{\rm Tr}(\mathbf{H}_l \mathbf{W}_m)+\sigma^2)  \leq {\rm Tr}(\mathbf{H}_l \mathbf{W}_k),\\\label{rankconst2.1}
 &\forall k,\:l=k,\ldots,N,  \\\label{rankconst2.4}
& \mathbf{W}_k \succcurlyeq 0,\\\label{rankconst2.5}
& \text{rank}(\mathbf{W}_k)=1.
\end{align}
\end{subequations}

{Note that} the rank-one constraint in \eqref{rankconst2.5} is non-convex.
To obtain a solution, the rank-one constraint is relaxed by exploiting the SDR approach.
Without the rank-one constraint, the following optimization problem is solved:
\begin{subequations}\label{rankone0}
\begin{align}
 \min_{\mathbf{W}_k\in \mathbb{C}^{M\times M}}  & \sum_{k=1}^{N} {\rm Tr}(\mathbf{W}_k) \\ \nonumber
s.t.\quad  & \gamma^{min}_k (\sum_{m=k+1}^{N}{\rm Tr}(\mathbf{H}_l \mathbf{W}_m)+\sigma^2)  \leq {\rm Tr}(\mathbf{H}_l \mathbf{W}_k),\\\label{rankconst02.1}
&\forall k,\:l=k,\ldots,N, \\\label{rankconst02.4}
& \mathbf{W}_k \succcurlyeq 0.
\end{align}
\end{subequations}

{Since \eqref{rankone0} is a standard semidefinite programming
(SDP), it can be efficiently solved
through convex optimization techniques.}
In general, if the solution of the relaxed problem in \eqref{rankone0} is a set of rank-one matrices $\mathbf{W}_k$, then it will be
also the optimal solution to the original problem in \eqref{rankone}. Otherwise, the randomization technique can be used to
generate a set of rank-one solutions \cite{6831610}.
{
The beamforming vector $\mathbf{w}_k$ can be obtained from a rank-one $\mathbf{W}_k$ solution, as $\mathbf{w}_k=\sqrt{\lambda_k} \mathbf{v}_k $
where $\lambda_k$ and $\mathbf{v}_k$ are the maximum eigenvalue and the corresponding eigenvector of $\mathbf{W}_k$,
respectively.}

\subsection{Complexity Analysis}
{In this paper, we have developed two approaches to transform the original non-convex optimization problem to a convex one.
In the SDR approach, the optimization problem is reformulated in SDP form by relaxing the non-convex rank one constraint. The optimal solution of the original
problem can be obtained from this simple SDR method if it yields rank-one solutions. On the other hand, it is possible in some cases for the solution of
the relaxed problem to turn out not to be rank-one. In this case, the proposed Taylor series approximation can be employed to convexify the original problem,
resulting in a suboptimal solution. We analyze the complexity of the proposed algorithms by evaluating the computational complexity of each problem based on the complexity of the interior point methods \cite{boydSOC,bookSDP}. 
This complexity can be defined by quantifying the required number of arithmetic operations in the worst-case at each iteration and
the required number of iterations to achieve the solutions with a certain accuracy. We define the computational complexity for each algorithm as follows.}

1) In the first scheme, the beamformer design in the power minimization problem is formulated into an second-order cone program (SOCP) in problem \eqref{taylor}.
Therefore, the worst case complexity is determined by the SOCP in each step. It is well known that for general interior-point methods the complexity of
the SOCP depends upon the number of constraints, variables and the dimension of each SOC constraint.  The total number of constraints in the formulation
of \eqref{taylor} is $0.5N^2+1.5N$. Therefore, the number of iterations needed to converge with $\varepsilon$ solution accuracy at the termination of the algorithm is
$O ( \sqrt{0.5N^2+1.5N}  \log⁡ \frac{1}{\varepsilon}  )$ \cite{boydSOC}.
Each iteration requires at most
$O \big((MN)^2 (0.33N^3+0.5N^2+1.16N+1)\big)$ arithmetic operations to solve the SOCP where MN and $0.33N^3+0.5N^2+1.16N+1$ are the number of optimization variables
and the total dimension of the SOC constraints in \eqref{taylor}.

2) The second scheme is a standard SDP. In this approach, the algorithm finds an $\varepsilon$-optimal solution for the semidefinite
problem with an $n$ dimensional semidefinite cone in at most $O ( \sqrt{n}  \log⁡ \frac{1}{\varepsilon}  )$ iterations where $n=M^2$ in our problem in \eqref{rankone0}.
Each iteration requires at most $O( m n^3  + m^2  n^2+ m^3 )$  arithmetic operations to solve
the SDP where $m$ denotes the number of semidefinite constraints \cite{bookSDP}.
Thus, $O\big(0.5 N(N+1)M^6+ 0.25 N^2 {(N+1)}^2 M^4+ 0.125 N^3 {(N+1)}^3 \big)$ arithmetic operations are required in each iterations of solving the problem in \eqref{rankone0}.

In summary, the first scheme has a much better worst-case complexity than an SDR scheme. In contrast to the semidefinite
formulation, there is no need to introduce the additional matrices $\mathbf{W}_k$ for the first scheme and the resulting optimization involves significantly fewer variables. However, in the first scheme, we have to deal with an approximation which makes the solution suboptimal. On the other hand, the SDR method can yield the optimal solution if it given a set of rank-one matrices which eliminates the need for the iterative approach as in the Taylor series approximation scheme. Note that the first scheme requires an iterative process, however, as seen in Fig. \ref{iteration2}, this approach converges with a small number of iterations which does not have significant impact on the order of the complexity of the proposed algorithm.

\section{Max-Min Fairness Problem} \label{sectionmaxmin}
In this section, we investigate a max-min fairness problem for the NOMA downlink system.
Since the users' rates in the NOMA scheme can be managed more flexibly,
it may be more appropriate {to provide a uniform user experience in terms of achieved throughput.
To balance the rate between different users in the network,
the max-min fairness approach is an appropriate criterion where the minimum achievable
rate of the users can be maximized} for a given total power constraint. The corresponding max-min fairness problem can be formulated as follows:
\begin{subequations}\label{maxproblem}
\begin{align}
  \max_{\mathbf{w}_k\in \mathbb{C}^{M\times 1}} & r, \\ \nonumber
 s.t.  & |\mathbf{h}^H_k \mathbf{w}_1|^2\geq\ldots\geq |\mathbf{h}^H_k \mathbf{w}_{k-1}|^2\geq|\mathbf{h}^H_k \mathbf{w}_k|^2 \\\label{maxconst1.2}
 &\qquad\qquad \geq|\mathbf{h}^H_k \mathbf{w}_{k+1}|^2\geq\ldots\geq|\mathbf{h}^H_k \mathbf{w}_N|^2, \forall k.\\\label{maxconst1.1}
  & \sum_{k=1}^N \| \mathbf{w}_k\|^2_2 \leq P^{max},
\end{align}
\end{subequations}
where $r = \min_{k} R_k$ is given in \eqref{SINR} and the constraint in \eqref{maxconst1.1} represents the maximum available total transmit power, i.e., $P^{max}$.
It is difficult to realize the optimal solution for this max-min problem due to {its} non-convex nature.
Hence, we first transform the problem into a convex one, then utilize a low-complexity polynomial algorithm to find an optimal solution.

\textit{Lemma 2:}
{This max-min fairness problem is quasi-concave and can be solved through a bisection search.}
\begin{IEEEproof}
{A maximization optimization problem is quasi-concave when the objective function is quasi-concave and the constraints are convex. The constraint \eqref{maxconst1.1} is convex, however the constraint \eqref{maxconst1.2} can be converted to a convex one
by using the methods provided in Section III.
Based on the quasi-convex definition \cite{7831468} 
a function $f$ is called quasi-convex if its domain and all its sublevel sets $S_{\alpha}=\{x\in dom f | f(x)\leq \alpha \}$, for $\alpha\in\mathbb{R}_+$, are convex. A function is quasi-concave if $-f$ is quasi-convex, i.e., every super level set $\{x\in dom f | f(x) \geq \alpha \}$ is convex. Clearly, for our objective function in \eqref{maxproblem}, $\min R_k (\mathbf{w})$,
to be quasi-concave, all its super level sets must be convex, i.e., $S_{\alpha}=\{ \min R_k (\mathbf{w}) \geq \alpha\}$,
which represents the set $\mathbf{w}=\{\mathbf{w}_1,\cdots, \mathbf{w}_N\}$ that makes the objective function
greater than a specific threshold, $\alpha$. Since there is the min operator, it can be rewritten as $S_{\alpha}=\{ R_k \geq \alpha, \forall k\}$ and the constraints $R_k \geq \alpha, \forall k$ can be obtained as
\begin{align} \nonumber
  &(2^{\alpha}-1)(\Sigma_{m=k+1}^{N}|\mathbf{h}_l^H \mathbf{w}_{m}|^2+\sigma^2)  \leq |\mathbf{h}_l^H \mathbf{w}_{k}|^2,\\ \label{maxmin}
 & \quad\qquad\qquad\qquad\qquad\qquad\qquad \forall k,~l=k,k+1,\ldots,N.
\end{align}}

{By employing the Taylor series approximation presented in Section III, the constraint in \eqref{maxconst1.2} and \eqref{maxmin}
can be transformed into a convex one, which completes the proof.}
\end{IEEEproof}

{Let us start with some transformations to simplify the constraints. From the inequalities in \eqref{maxconst1.2}, it holds that}

\begin{align}\nonumber
&\left\{\begin{array}{l l} |\mathbf{h}^H_k \mathbf{w}_2|^2\leq |\mathbf{h}^H_k \mathbf{w}_1|^2,\\
\vdots \\
|\mathbf{h}^H_k \mathbf{w}_{k+1}|^2\leq \min_{m\in[1,k]} |\mathbf{h}^H_k \mathbf{w}_m|^2,\\ 
\vdots \\
|\mathbf{h}^H_k \mathbf{w}_{N}|^2\leq \min_{m\in[1,N-1]} |\mathbf{h}^H_k \mathbf{w}_m|^2,
 \end{array}
\right. \\ \nonumber
&\Leftrightarrow |\mathbf{h}^H_k \mathbf{w}_{n}|^2 \leq |\mathbf{h}^H_k \mathbf{w}_{m}|^2, \\ \label{const2}& \qquad \forall k,\:n=2,\ldots,N,\:m=1,\ldots,n-1. 
\end{align}

{After these simplifications, Lemma 1 in Section III can be employed to convexify \eqref{const2} as
\begin{align}\nonumber
&|\mathbf{h}^H_k \mathbf{w}_{n}|^2 \leq g_k(\mathbf{w}_m, \mathbf{w}_m^t), ~~~\forall k,\:n=2,\ldots,N,\\ \label{newconst2}
&\qquad\qquad\qquad\qquad\qquad\qquad m=1,\ldots,n-1. 
\end{align}
where $g_k(\mathbf{w}_m, \mathbf{w}_m^t)$ is the Taylor series approximation of the term $|\mathbf{h}^H_k \mathbf{w}_{m}|^2$ around $\mathbf{w}_m^t$ in the $t^{th}$ iteration. }

{Similarly, the equivalent convex formulation for \eqref{maxmin} can be reformulated as
\begin{align} \nonumber
  &(2^{\alpha}-1)(\Sigma_{m=k+1}^{N}|\mathbf{h}_l^H \mathbf{w}_{m}|^2+\sigma^2)  \leq g_l(\mathbf{w}_k, \mathbf{w}_k^t),  \\ \label{maxmin2}
 & \quad\qquad\qquad\qquad\qquad\qquad \forall k,~l=k,k+1,\ldots,N.
\end{align}
where $g_l(\mathbf{w}_k, \mathbf{w}_k^t)$ is the Taylor series approximation of the term $|\mathbf{h}_l^H \mathbf{w}_{k}|^2$ around $\mathbf{w}_k^t$ in the $t^{th}$ iteration.}

\begin{table}[t] \caption{Bisection method}\label{bisection}
\centering
\begin{small}
\begin{tabular} {|l|}
\hline
\textbf{Algorithm 2} ~Proposed Algorithm for solving problem \eqref{maxproblem}\\
\hline
1. \textbf{Initialization:}~ Set $t_{min}=0,{ t_{max}=\log_2\big(1+\frac{P^{max}|\mathbf{h}_N|^2}{\sigma^2}\big)}$,\\ 
2. \textbf{repeat}\\
3. \qquad Set $t=(t_{max}+t_{min})/2$ and solve \eqref{maxproblem2} to obtain $\mathbf{w}_0$ \\
4. \qquad \textbf{if}  \eqref{maxconst1.1} is satisfied \textbf{then}\\
5. \qquad \qquad Set $t_{min}=t; \mathbf{w}^*=\mathbf{w}_0 ; R^*=t$\\
6.  \qquad \textbf{else}\\
7.  \qquad \qquad $t_{max}=t$\\
8. \textbf{until} $(t_{max}-t_{min}\leq\varepsilon)$.\\
\hline
\end{tabular}
\end{small}
\end{table}

In order to solve this problem through a bisection method,
assume that $R^*$ denotes  the optimal value of the objective function of the problem in \eqref{maxproblem}. 
{ For a given threshold $\alpha$, if there exists a set of $\mathbf{w}_0=\{\mathbf{w}_1,\cdots, \mathbf{w}_N\}$ that satisfies the constraints
\eqref{maxconst1.1},\eqref{newconst2} and \eqref{maxmin2}, then $R^*\geq \alpha$, otherwise $R^*\leq \alpha$. Equivalently, the following problem can be solved}
\begin{subequations}\label{maxproblem2}
\begin{eqnarray}
\min_{\mathbf{w}_k\in \mathbb{C}^{M\times 1}} && \sum_{k=1}^{N} \| \mathbf{w}_k\|^2_2, \\\label{max1.1}
  s.t.  && \eqref{newconst2} ~\text{and} ~\eqref{maxmin2},
\end{eqnarray}
\end{subequations}
{{and determined whether the solution satisfies the total power constraint  $\sum_{k=1}^N \| \mathbf{w}_k\|^2_2 \leq P^{max}$. By appropriately choosing $\alpha$ through a bisection method, the solution of \eqref{maxproblem} can be obtained
by solving a sequence of feasibility problems of
\eqref{maxproblem2}.} Table \ref{bisection} presents the proposed
bisection method for realizing the solution for the problem in \eqref{maxproblem}.}

\section{Robust Power Minimization}
{In previous sections, it has been assumed
that perfect CSI is available at the transmitter, which might be difficult under practical conditions
due to estimation and quantization errors. In this section, to circumvent
the inevitable channel uncertainties, we study a robust design for the problem
considered in Section III.
In particular, we consider the robust optimization problem with the outage probability constraints by incorporating channel uncertainties.}
It is assumed {that an} imperfect estimate of the channel covariance matrix is available at the BS.
Let $\mathbf{\hat{C}}_k=\mathbb{E}(\mathbf{\hat{h}}_k {\mathbf{\hat{h}}_k}^H) \in \mathbb{C}^{M\times M}$ denote the \mbox{estimated}
channel covariance matrix of $U_k$ and the corresponding uncertainty matrix is denoted by
\mbox{$\mathbf{\Delta}_k \in \mathbb{C}^{M\times M} $}. 
The $ij^{th}$ entry of $\mathbf{\Delta}_k$ is independently and identically
distributed as $[\mathbf{\Delta}_k]_{ij} \sim \mathcal{CN}(0, \sigma^2_{ij})$.
Hence, the actual channel covariance matrix can be modelled as
\begin{equation}
  \mathbf{C}_k=\mathbf{\hat{C}}_k+\mathbf{\Delta}_k, \qquad \forall k .
\end{equation}

Based on the SIC approach in Section II,
{the SINR of the signal intended for the $k^{th}$ user at the $l^{th}$ user can be written as}
\begin{equation}\label{robust}
 \text{SINR}^l_k=\dfrac{\mathbf{w}_k^H(\mathbf{\hat{C}}_l+\mathbf{\Delta}_l)\mathbf{w}_k}{
\sum_{m=k+1}^{N} \mathbf{w}_m^H(\mathbf{\hat{C}}_l+\mathbf{\Delta}_l)\mathbf{w}_m+\sigma^2}.
\end{equation}

In the sequel, we reformulate the optimization problem
by taking into account the channel uncertainties as

\begin{subequations}\label{robustproblem}
\begin{align}
&\min_{\mathbf{w}_k\in \mathbb{C}^{M\times 1}} \sum_{k=1}^N  \| \mathbf{w}_k\|^2_2, \\ \nonumber
&s.t. \!\min_{l \in \{k, k+1, \ldots, N\}} \dfrac{\mathbf{w}_k^H(\mathbf{\hat{C}}_l+\mathbf{\Delta}_l)\mathbf{w}_k}{
\sum_{m=k+1}^{N} \mathbf{w}_m^H(\mathbf{\hat{C}}_l+\mathbf{\Delta}_l)\mathbf{w}_m+\sigma^2}\!\geq\!\!
\gamma^{min}_k, \\ \label{rob1.1}
& \qquad\qquad\qquad\qquad\qquad\qquad\qquad\qquad\qquad\qquad\: \forall k,
\end{align}
\end{subequations}
where $\gamma^{min}_k$ is the minimum required SINR at the user $U_k$.
Then, \eqref{rob1.1} can be rewritten as
\begin{align}\nonumber
&\gamma^{min}_k\big(\!\!\!\sum_{m=k+1}^{N} \mathbf{w}_m^H(\mathbf{\hat{C}}_{l}+\mathbf{\Delta}_{l})\mathbf{w}_m+\sigma^2\big)
\leq \mathbf{w}_k^H(\mathbf{\hat{C}}_{l}+\mathbf{\Delta}_{l})\mathbf{w}_k, \\
 & \qquad \qquad \qquad\qquad\qquad\qquad\qquad\quad \forall k,\:l=1,\ldots,N.
\end{align}

\subsection{Outage probability based robust design}
{In practical scenarios, the channel parameters are prone to error, hence
the robust beamforming design against statistical channel uncertainties is an important issue that needs to be addressed.
By applying the outage probability to \eqref{robustproblem}, the robust power minimization problem can be reformulated as}
\begin{subequations}\label{robustproblem2}
\begin{align}
&\min_{\mathbf{w}_k\in \mathbb{C}^{M\times 1}}\sum_{k=1}^ N \| \mathbf{w}_k\|^2_2, \\ \nonumber
&s.t.\;\:\, {\rm Pr}\Big(\gamma^{min}_k\big(\!\!\!\sum_{m=k+1}^{N} \mathbf{w}_m^H(\mathbf{\hat{C}}_{l}+
\mathbf{\Delta}_{l})\mathbf{w}_m+\sigma^2\big)
\leq \\\label{rob21.1}
&\qquad \mathbf{w}_k^H(\mathbf{\hat{C}}_{l}+\mathbf{\Delta}_{l})\mathbf{w}_k \Big)
\geq (1-\rho_k),\quad \forall k,\:l=1,\ldots,N,
\end{align}
\end{subequations}
{where $\rho_k\in (0, 1)$ is the outage probability at $U_{k}$. }In other words, the predefined probability of satisfying the required SINR
at the user $U_{k}$ is $(1-\rho_{k})$.
The above robust problem in \eqref{robustproblem2} is NP-hard and cannot be solved directly. In order to determine the solution for
this robust problem, we introduce a new matrix variable $\mathbf{W}_k=\mathbf{w}_k \mathbf{w}_k^H$ and utilize a procedure to convert probabilistic constraints into a
tractable form.

\textit{Lemma 3:}
The original robust power minimization problem in
\eqref{robustproblem2} can be reformulated as
\begin{subequations}\label{finalrobustproblem}
\begin{align}
\min_{\mathbf{W}_k\in \mathbb{C}^{M\times M}} & \sum_{k=1}^N {\rm Tr} (\mathbf{W}_k), \\\label{finalrob1.1}
s.t. & \mathbf{C}_{kl} \succeq 0,\;\; \forall k,~l=k,k+1,\ldots,N,\\ \label{finalrob1.3}
& \mathbf{W}_k \succeq 0 \\ \label{finalrob1.4}
& \text{rank}(\mathbf{W}_k) = 1.
\end{align}
\end{subequations}
{where $\mathbf{C}_{kl}$ is defined as \eqref{con1} in Appendix \ref{App.C}.}

\begin{IEEEproof}
Please refer to Appendix \ref{App.C}.
\end{IEEEproof}

\vspace{0.5cm}
The constraints in \eqref{finalrob1.1} and \eqref{finalrob1.3} are semidefinite in terms of $\mathbf{W}_k$.
{Therefore, the optimization problem in \eqref{finalrobustproblem}
is a standard SDP without the non-convex rank-one constraint in \eqref{finalrob1.4}.
This optimization problem can be solved through relaxing the
non-convex rank-one constraint. In general, if the solution of the relaxed problem is a set of rank-one matrices $\mathbf{W}_k$, then it will be
also the optimal solution to the original problem in \eqref{finalrobustproblem}. Otherwise, the randomization technique can be used to
generate a set of rank-one solutions \cite{6831610}.
The beamforming vector $\mathbf{w}_k$ can be determined through extracting the maximum eigenvalue and the corresponding eigenvector of $\mathbf{W}_k$.}

\begin{figure}[t!]
  \begin{center}
    \includegraphics[scale=0.28]{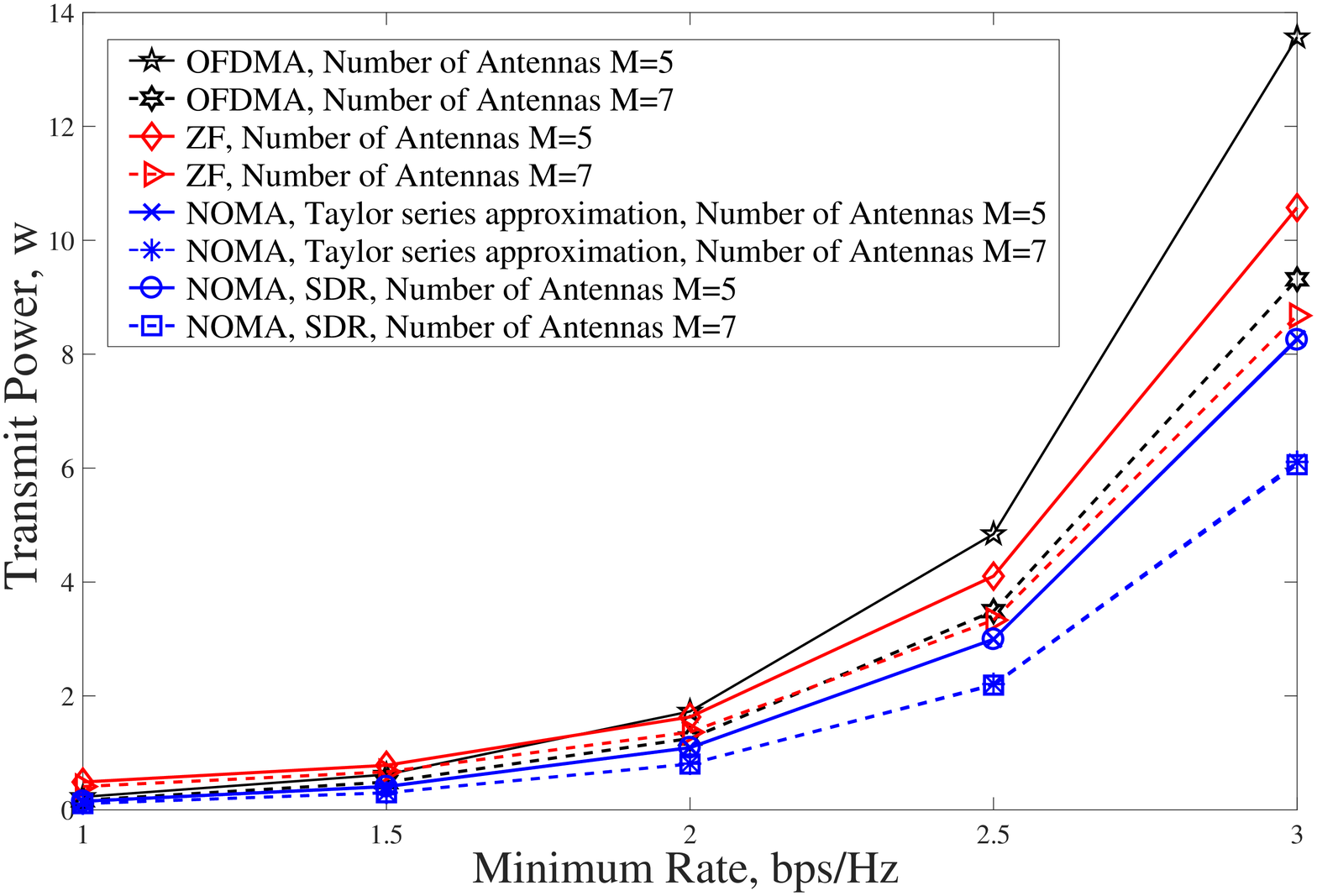} 
    \vspace{-0.5cm}
    \caption{The required total transmit power to achieve different target rates for $3$ users in NOMA, ZF and OFDMA schemes.}
    \label{PR}
\vspace{0.3cm}
    \includegraphics[scale=0.28]{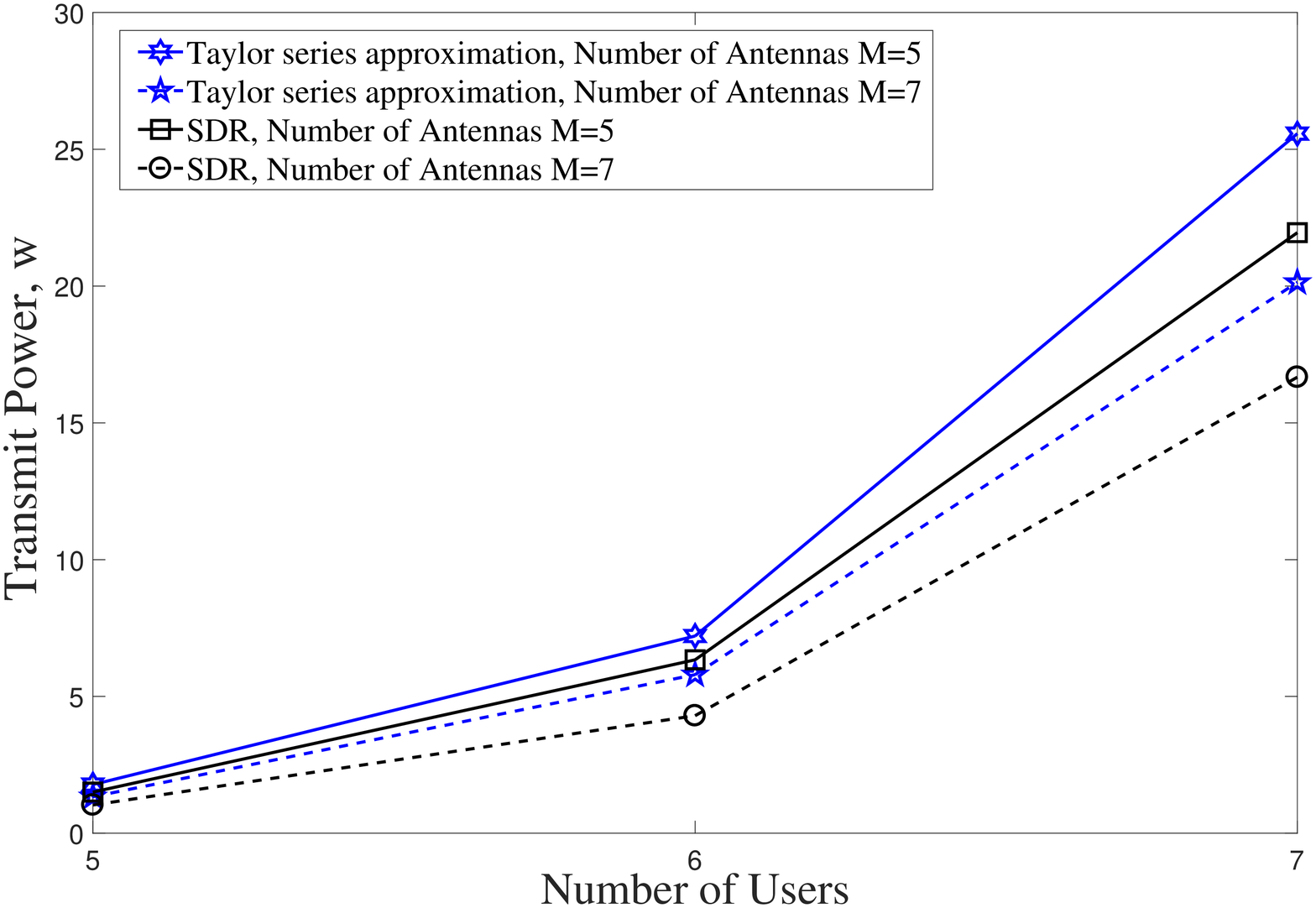} 
    \vspace{-0.5cm}
    \caption{The required total transmit power to achieve $R_k^{min}=2$ bps/Hz for different numbers of users by using Taylor series approximation and SDR methods.}
    \vspace{-0.5cm}
    \label{PN}
  \end{center}
\end{figure}

\section{Simulation Results}
In this section, the performance of the proposed beamforming designs for the NOMA scheme
 is evaluated through numerical simulations.
{We consider a single cell downlink transmission,
where a multi-antenna BS serves single-antenna
users which are uniformly distributed within the circle with
a radius of $50$ meters around the BS, but no closer than $1$ meter.
The small-scale fading of the channels is assumed to be
Rayleigh fading which represents an isotropic scattering environment. The large-scale fading effect is modelled by
${d_k}^{-\beta}$ to incorporate the path-loss effects where $d_k$ is the
distance between $U_k$ and the BS, measured in meters and
$\beta$ is the path-loss exponent. Hence, the channel coefficients between the BS and user $U_k$ are generated using $\mathbf{h}_k=\chi_k  \sqrt{{d_k}^{-\beta}}$ where $\chi_k \sim \mathcal{CN} (0,\mathbf{I})$   and $\beta= 3.8$ \cite{7446365}.
It should be noted here that in
simulations the user distances are fixed and the average is taken
over the fast fading component of the channel vectors.}
It is assumed
that the noise variance at each user is 0.01 ($\sigma^2=0.01$) and the
target rates for all users are the same.
The term non-robust scheme refers to the scheme where the BS has imperfect
CSI without any information on the channel uncertainties and
the beamforming vectors are designed based on imperfect CSI
without incorporating channel uncertainty information.
In addition, for the imperfect-CSI case the variance of each entry (i.e., $[\mathbf{\Delta}_{k}]_{ij}$ ) of the error covariance matrix $\mathbf{\Delta}_{k}$
and the predefined outage probability ($\rho_{k}$) of the required QoS constraints are set
to $0.005$ and $0.1$, respectively.
These numerical results are obtained by averaging over different 1000
random channels.

\begin{figure}[t!]
  \begin{center}
    \includegraphics[scale=0.28]{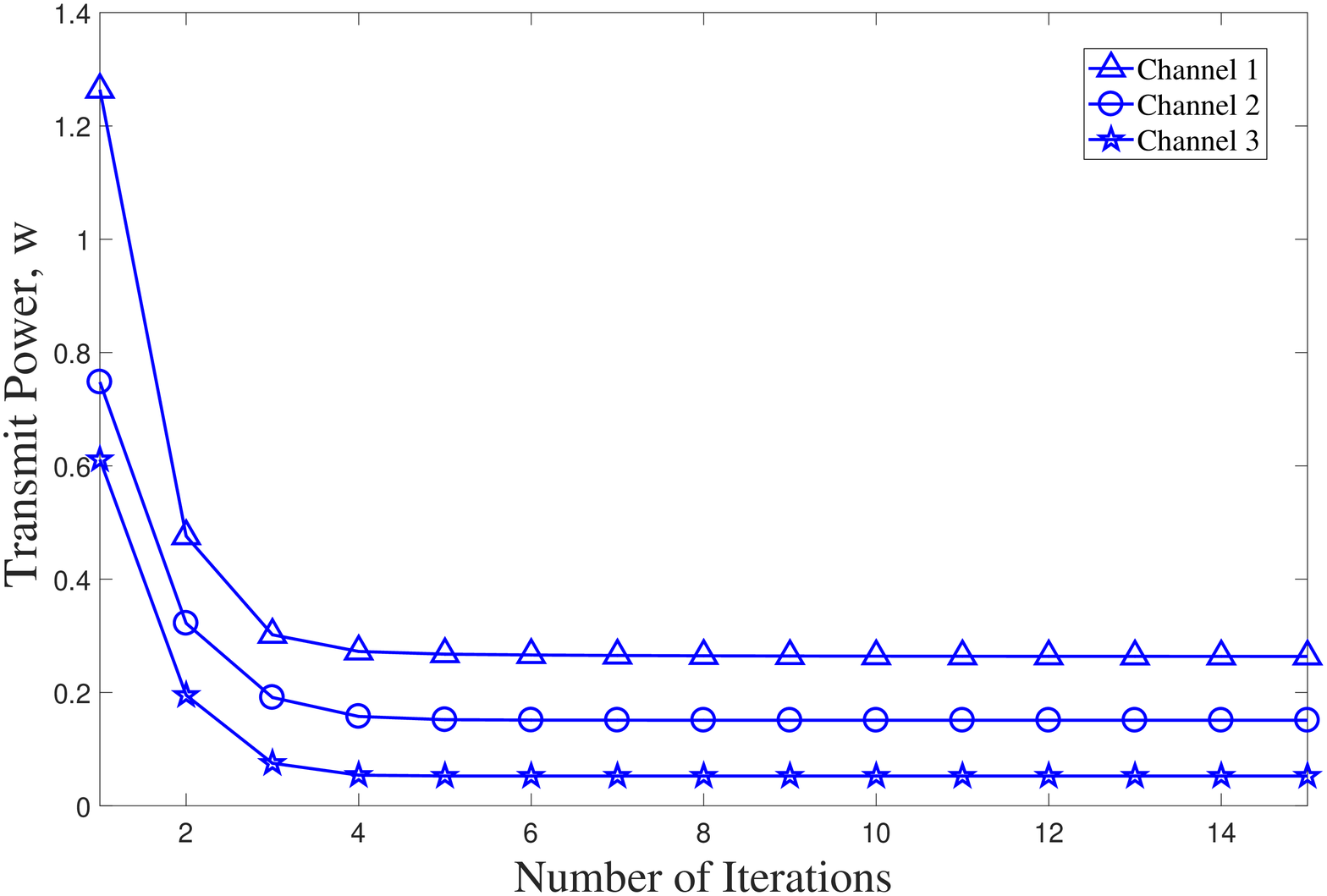} 
     \vspace{-0.5cm}
    \caption{The convergence of the algorithm in Table \ref{tayloralgorithm} for different set of channels. Number of users$=3$, ~Number of antennas$=6$, ~Target rate$=1$. }
    \label{iteration2}
    \includegraphics[scale=0.35]{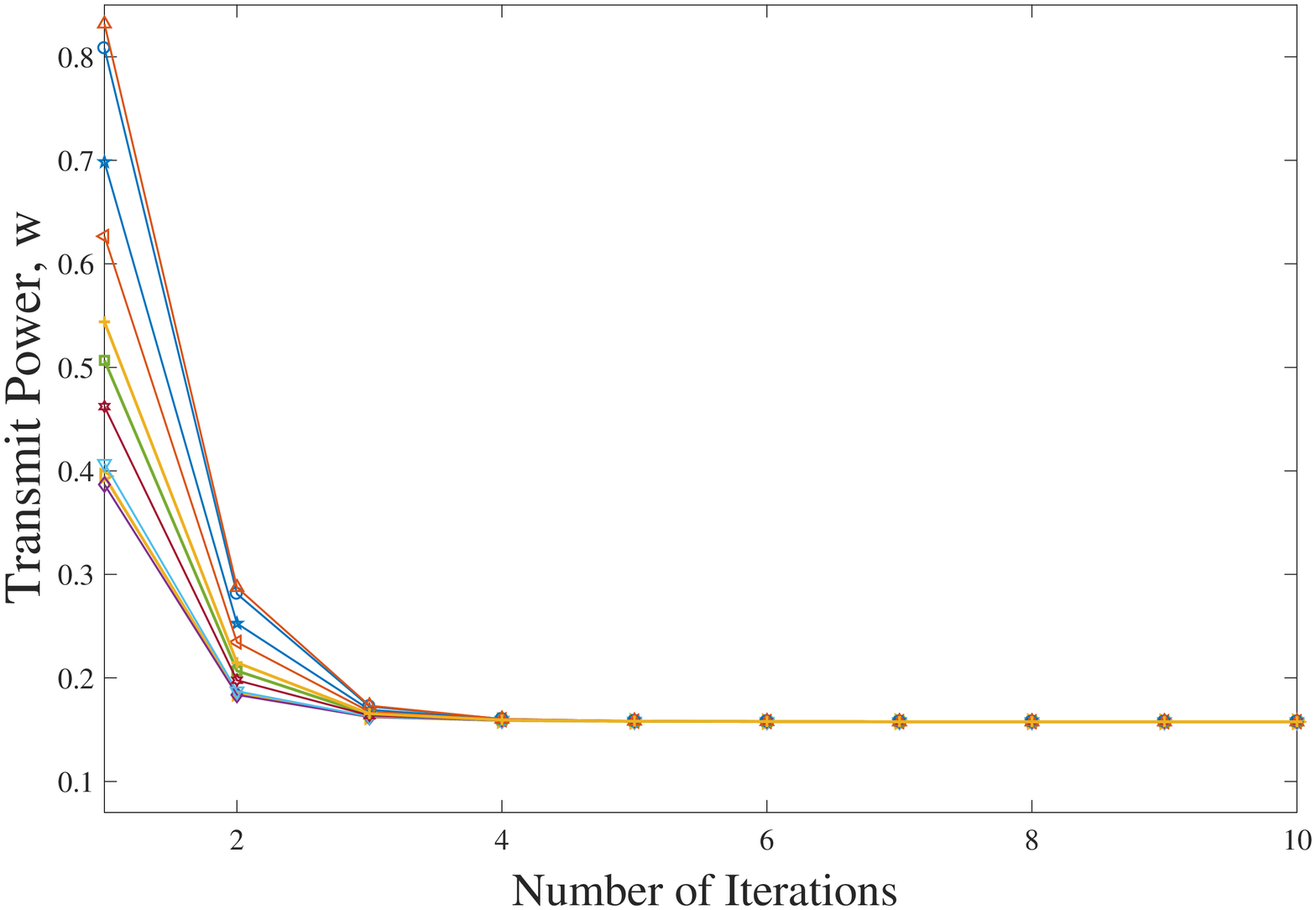} 
     \vspace{-0.5cm}
    \caption{The convergence of the algorithm based on Taylor series approximation for different initializations. Number of users$=3$, ~Number of antennas$=5$, ~Target rate$=1$.}
    \vspace{-0.5cm}
    \label{iteration}
  \end{center}
\end{figure}

\subsection{Power minimization and max-min fairness designs}
First, we evaluate the required total transmit power for both power minimization approaches (i.e., Taylor series approximation and SDR) with different system parameters. As can be seen from the figures, the results of these two methods are almost identical.
The required total transmit power against different target rates is presented in Fig. \ref{PR} for the NOMA and OMA schemes with different numbers of transmit antennas. By increasing the minimum required
rate at each user, the BS requires more power to satisfy the target rate requirements. For a given target rate,
the required total transmit power can be reduced by employing more antennas at the transmitter.
As shown in Fig. \ref{PR}, for a specific rate requirement, the conventional OMA technique
consumes more transmit power than the NOMA scheme.
This demonstrates that the NOMA scheme outperforms the conventional OMA in terms of energy efficiency.

In Fig. \ref{PN}, the required total transmit power for different {numbers} of users with different numbers of transmit
antennas is obtained.
As the number of antennas increases, the required transmit power decreases due to the spatial diversity gain.
However, the BS requires more transmit power as the number of users increases.
{As shown in Fig. \ref{PN}, both schemes, Taylor series approximation and SDR show a similar performance for a few users due to the small number of approximated terms in the Taylor series approximation. However, the number of approximated terms increases with the number of users. As a result, the performance gap between these two schemes increases and SDR outperforms the Taylor series approximation scheme in terms of required transmit power. The reason is that SDR can provide the optimal solution given that the solution is rank one whereas the other scheme relies on the Taylor series approximation which might lead to a suboptimal solution.}
Fig. \ref{iteration2} depicts the convergence of the algorithm provided in Table \ref{tayloralgorithm} in terms of transmit power.
{As shown, this approach converges with a small number of iterations (most of the time with $3$ iterations), which does not have a significant impact on the order of the complexity of the proposed algorithm. Moreover, we have numerically evaluated the impact of the initialization of the algorithm on the convergence of the Taylor series approximation. As shown in Fig. \ref{iteration}, the Taylor series approximation method converges to the same solution with different initializations.}

Table \ref{table1} is provided to compare the required transmit power for each user and the total transmit power obtained through
the Taylor series approximation and SDR approaches.
{As evidenced by these results, there {is no} significant difference between the two proposed approaches.}

\begin{figure}[t!]
  \begin{center}
    \includegraphics[scale=0.28]{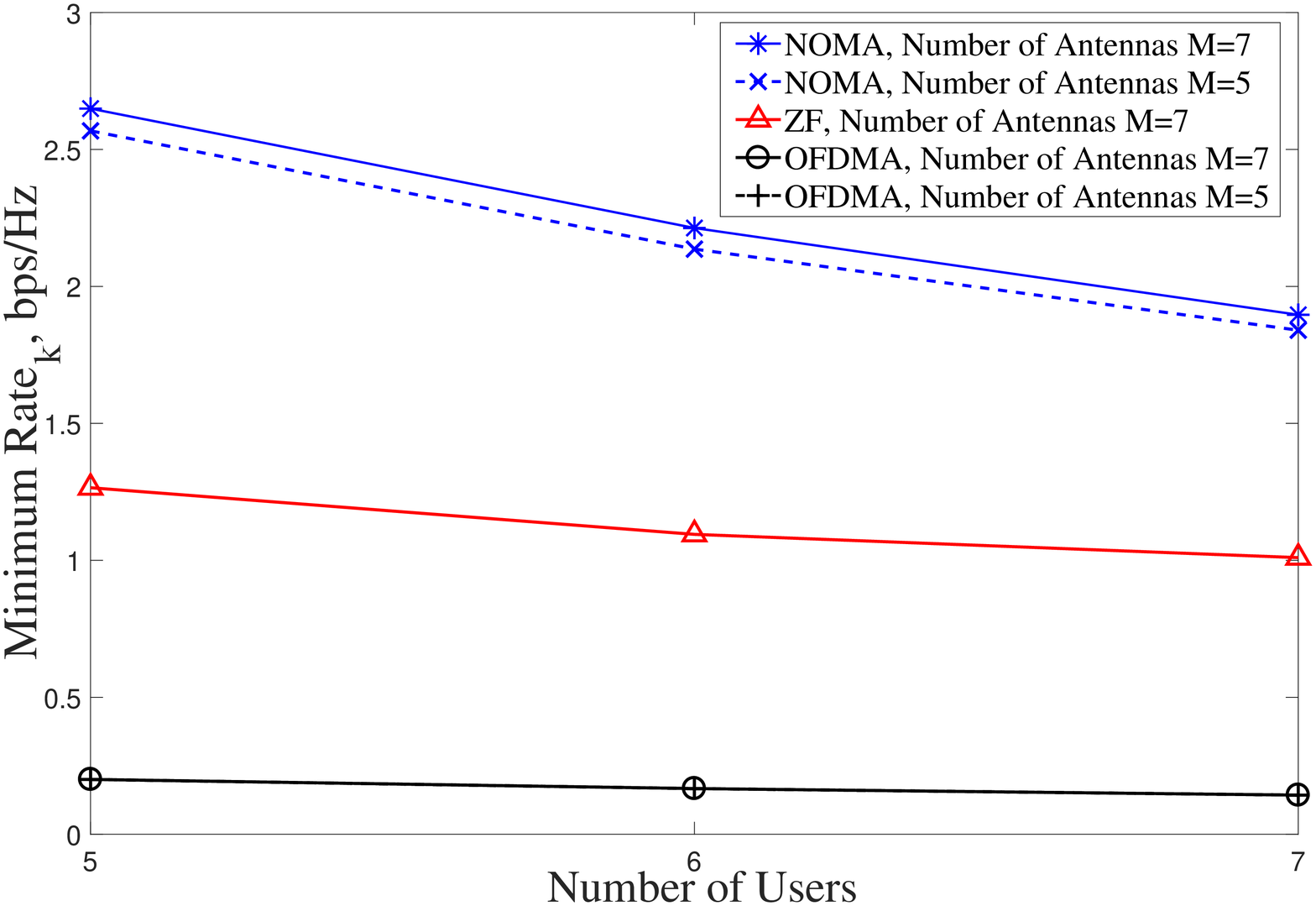}
     \vspace{-0.3cm}
    \caption{The minimum achieved rate for different numbers of users with $P^{max}=10 w$ in NOMA, ZF and OFDMA schemes.}
    \vspace{-0.5cm}
    \label{RN}
    \vspace{0.5cm}
    \includegraphics[scale=0.28]{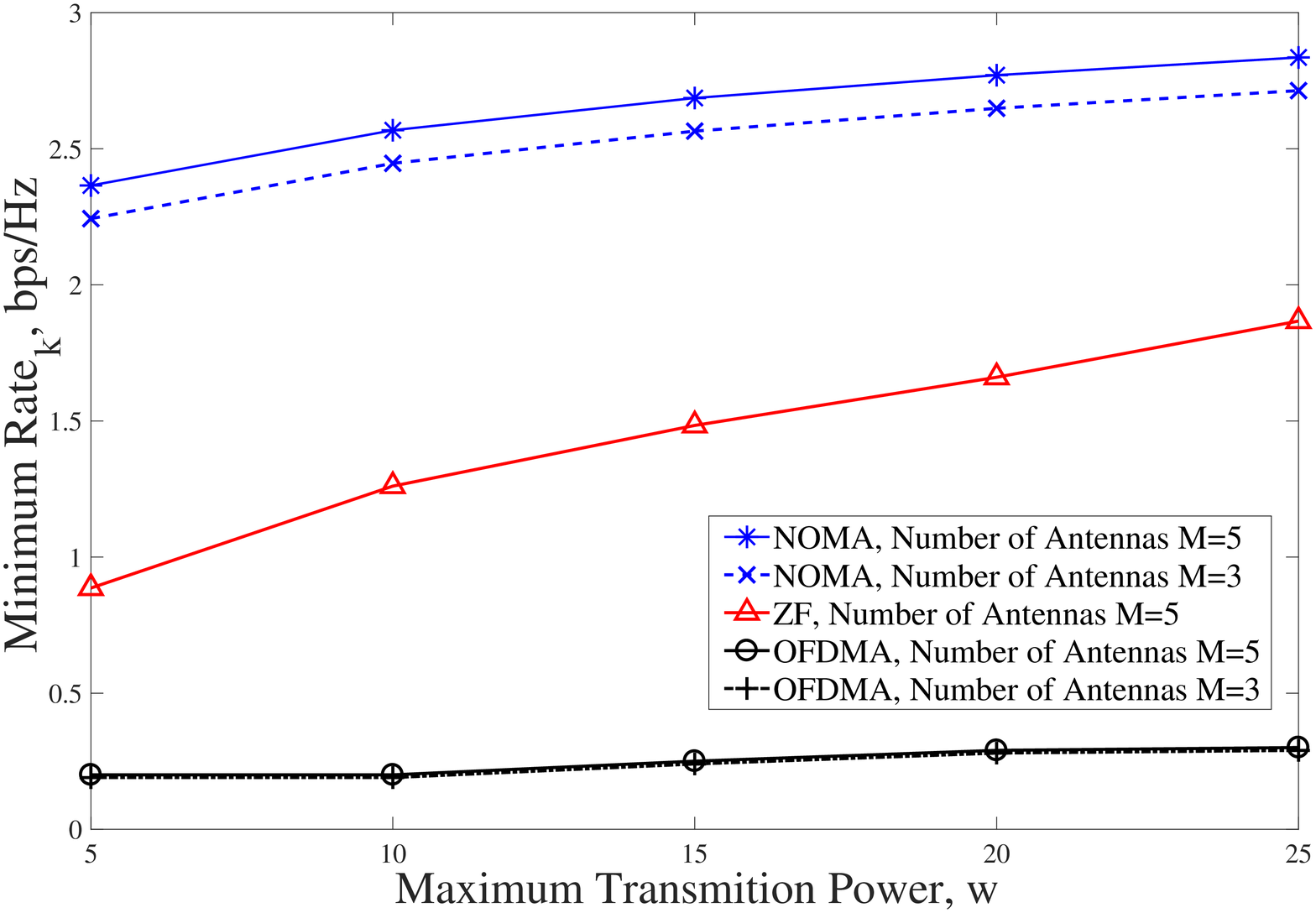}
     \vspace{-0.3cm}
    \caption{The minimum achieved rate for $5$ users with different $P^{max}$ in NOMA, ZF and OFDMA schemes.}
    \vspace{-0.5cm}
    \label{RP}
    \vspace{-0.2cm}
  \end{center}
\end{figure}

{Next, we study the performance of the max-min fairness design for both NOMA and OMA schemes.
The balanced rates maintaining fairness between users are demonstrated} in Fig. \ref{RN}  and Fig. \ref{RP}, respectively,
for different numbers of users and maximum available transmit power with different numbers of antennas.
As expected, reducing the numbers of users or increasing the maximum available transmit power threshold improves the achievable fairness rate. Since the fairness rate is a logarithmic function of power, as the power threshold increases the rate improvement is compressed.
These simulation results confirm that the QoS based
beamforming design satisfies the required rate constraints at each
user whereas the rates of the users are balanced in the fairness based approach.
 {As shown in Fig. \ref{RN}  and Fig. \ref{RP}, for a specific available power, the NOMA scheme achieves more rate than the conventional OMA technique.}

The power allocations and the balanced rates obtained by solving problem \eqref{maxproblem}
are provided for five different random channels in Table \ref{table2}.
{In order to validate the optimality of the proposed max-min fairness approach,
we compare these with the power allocations through the power minimization solution in Section III.
In particular, the balanced rates obtained through the fairness approach have been set as the target rates
in the power minimization approach for the same set of channels, and the corresponding power allocations
are obtained. As seen in Table \ref{table2} and Table \ref{table3}, both max-min fairness and power minimization approaches utilize the same
power allocations to achieve same rates at each user. This confirms the optimality of the proposed
max-min fairness based design as the power minimization approach is optimal for a given set of target rates.}

\subsection{Performance Study of Robust Design}
In this subsection, we study the impact of the proposed robust design on the achieved rate in comparison with the
non-robust scheme.
{The effect of error variances on the required transmit power is represented in Fig. \ref{delta}. It can be observed
from Fig. \ref{delta} that the total transmit power at the BS increases as
the errors in the CSI increase. }

We compare the performance of the robust and the non-robust scheme through the rate satisfaction ratio $\eta_k$,
which is defined as the ratio between the achieved rate and the target rate at the user $U_{k}$.
Hence, $\eta_k\geq1$ indicates that the rate requirement is satisfied at the user $U_{k}$.

Fig. \ref{ro} and Fig. \ref{nro} depict the histogram of the rate satisfaction ratio for the
robust and the non-robust schemes with the target rate, $R_{min}=3$, respectively.
{We also study the rate satisfaction ratio for the
robust and the non-robust OMA schemes in Fig. \ref{roma} and Fig. \ref{nroma}, respectively.}
It can be observed that in the robust design, rate constraint is satisfied in most cases and
in only $10\%$ of cases does the rate satisfaction ratio fall below one according to the outage probability requirement.
However, as evidenced by results
presented in Fig. \ref{nro} and \ref{nroma}, the non-robust design cannot satisfy the target rate requirement for approximately $50$ percent of the cases since it does not take into account any information regarding channel uncertainties.

\begin{figure}[t!]
  \begin{center}
    \includegraphics[scale=0.28]{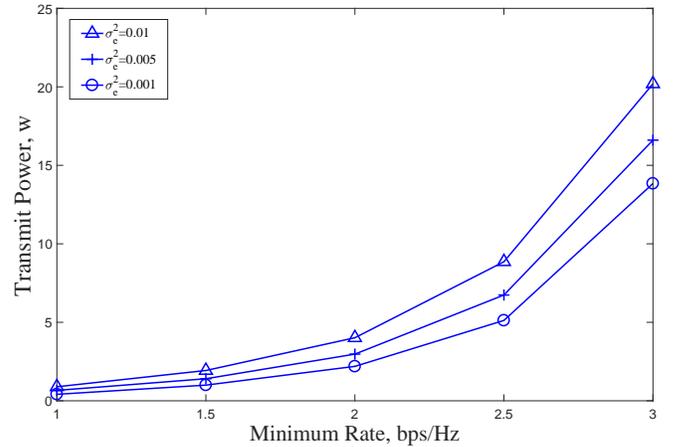} 
     \vspace{-0.3cm}
    \caption{The required total transmit power to achieve different target rates with different channel uncertainties at fix outages $\rho=0.1$. Number of users = Number of antennas = 3.}
    \vspace{-0.5cm}
    \label{delta}
  \end{center}
\end{figure}

\section{Conclusions}
In this paper, we have proposed different beamforming techniques for
NOMA based downlink transmission. In particular, these beamforming designs were developed through
a) a power minimization approach to achieve the required target rate at each user; b) a max-min fairness
approach to maintain user fairness in terms of the achieved rates and c) an outage probability based
robust approach to satisfy the target rates with a set of predefined probabilities. To tackle the
original non-convex problems, we have developed iterative algorithms by exploiting first order Taylor
series approximations and SDR techniques for the first two problems whereas the robust design was solved
through converting the non-convex constraints into a set of convex LMIs.
Simulation results were provided to validate the performance of the
proposed schemes in terms of the required transmit power and balanced rates.
These results confirm
that the proposed outage probability based robust approach
outperforms the non-robust scheme in terms of the achieved
rates and rate satisfaction ratio at each user. {These simulation results also demonstrate that NOMA can achieve superior performance in terms of system throughput compared to the traditional multiple access and can efficiently utilize the bandwidth resources.}
In this work, we developed novel resource allocation techniques to improve the system throughput and maintain user fairness, which address the issues associated with multiple access scheme in next generation wireless networks.

\begin{figure}[t!]
  \begin{center}
    \includegraphics[scale=0.28]{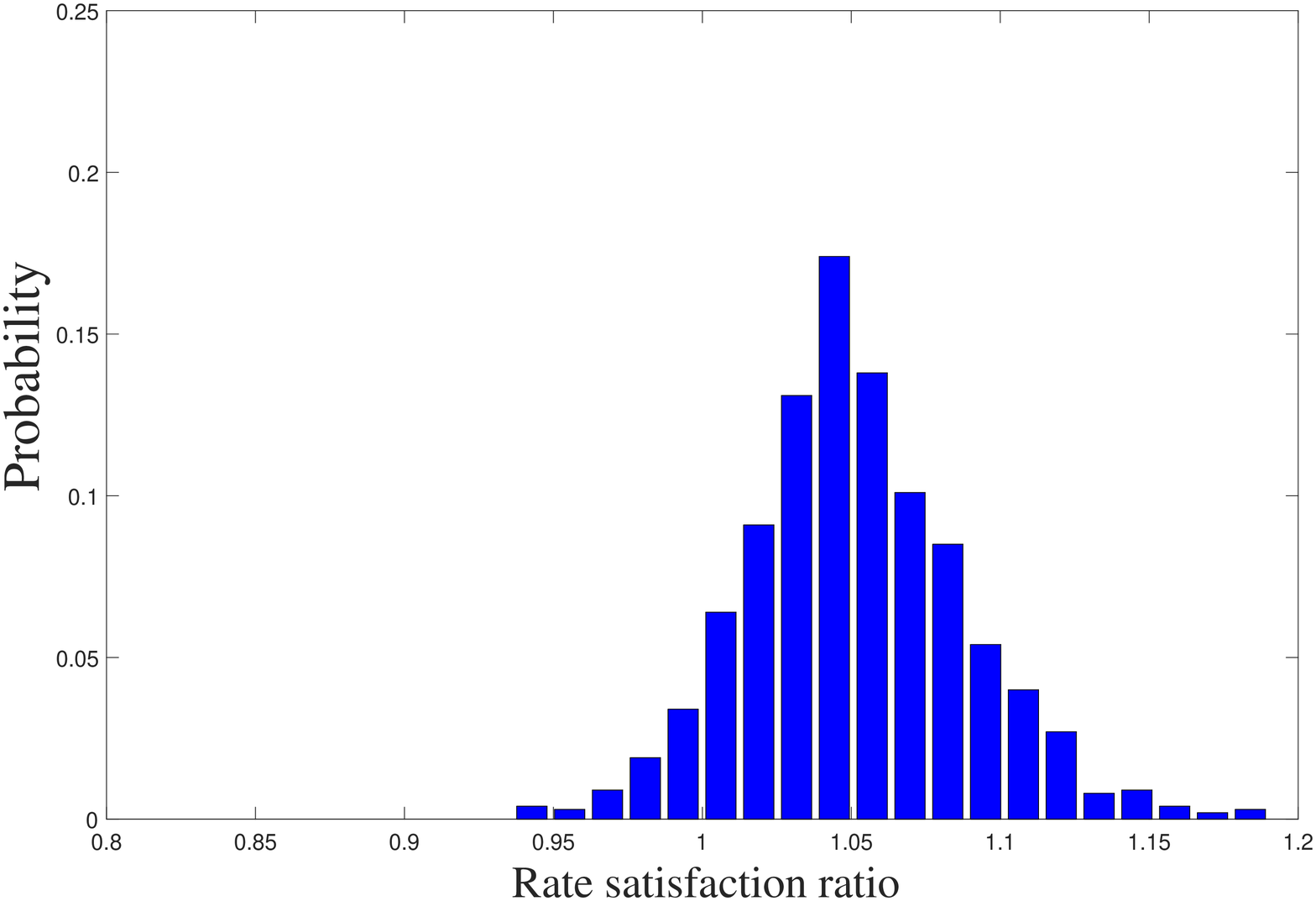} 
    \vspace{-0.6cm}
    \caption{Histogram for rate satisfaction ratio, i.e., $\eta_k$, for $R_{min}=3 bps/Hz$ in the robust NOMA scheme. } \vspace{-0.3cm}
    \label{ro}
    \vspace{0.5cm}
    \includegraphics[scale=0.28]{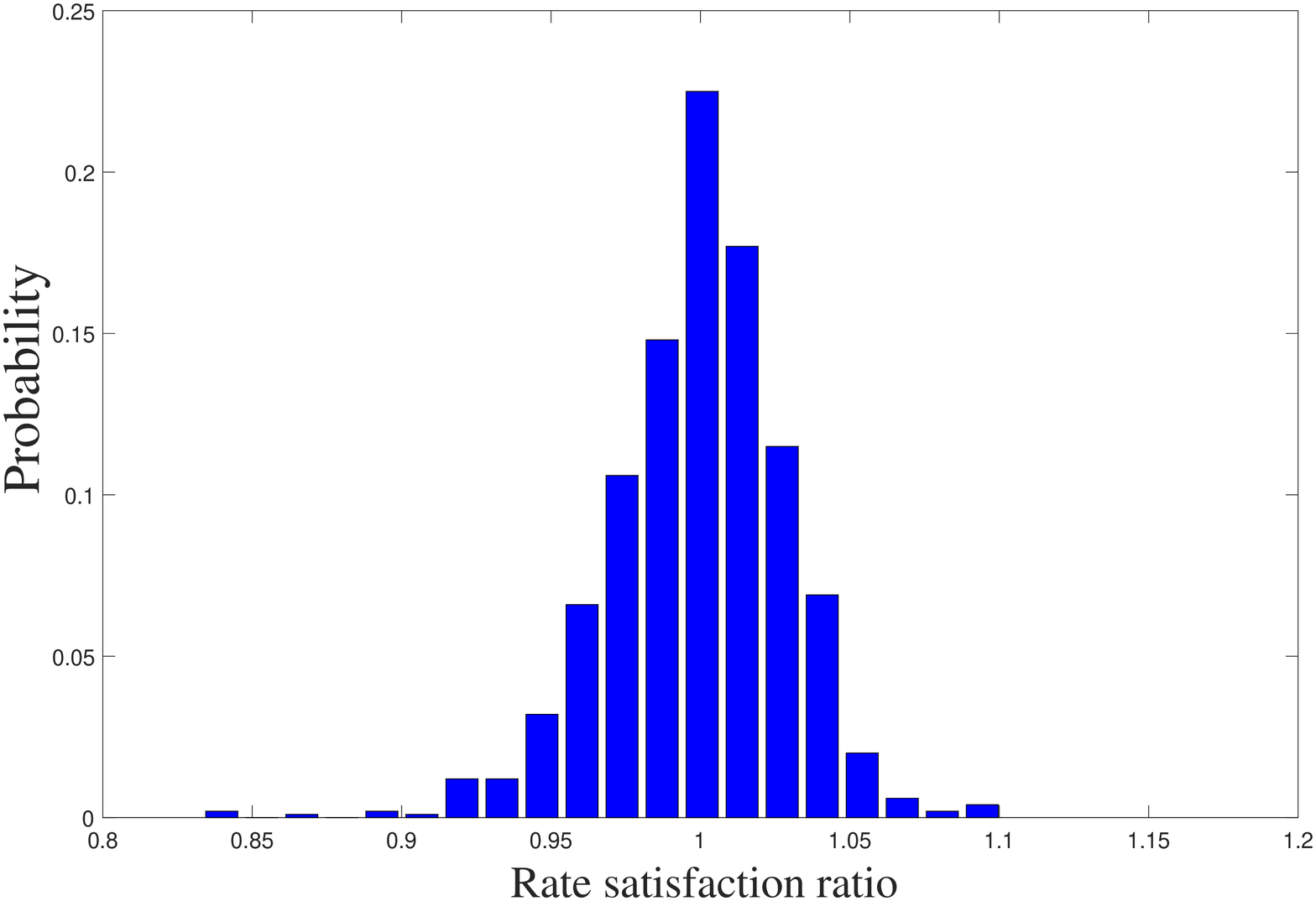} 
    \vspace{-0.6cm}
    \caption{Histogram for rate satisfaction ratio, i.e., $\eta_k$,  for $R_{min}=3 bps/Hz$ in the non-robust NOMA scheme.}\vspace{-0.3cm}
    \label{nro}
  \end{center}
\end{figure}

\begin{figure}[t!]
  \begin{center}
    \includegraphics[scale=0.28]{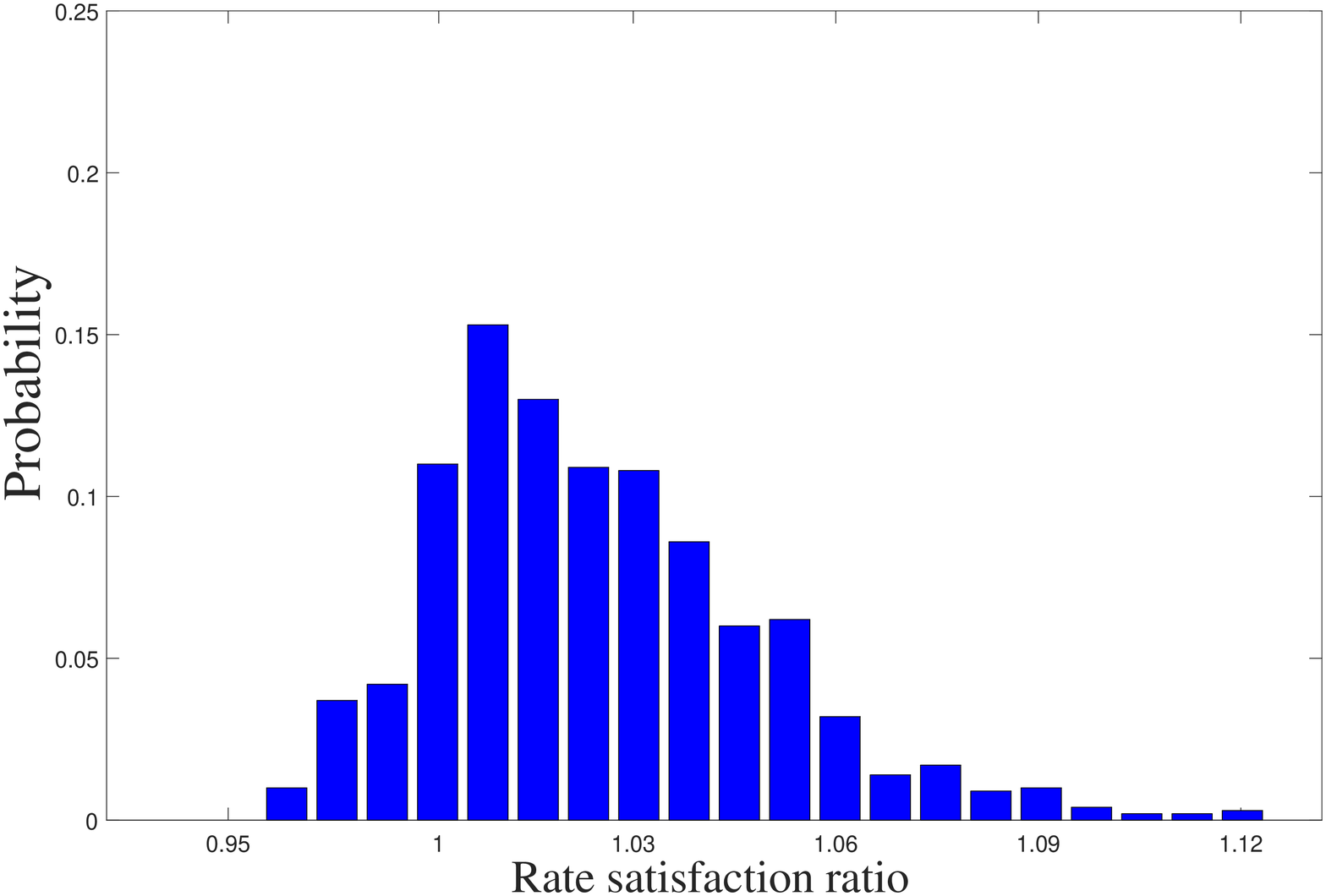} 
    \vspace{-0.6cm}
    \caption{Histogram for rate satisfaction ratio, i.e., $\eta_k$, for $R_{min}=3 bps/Hz$ in the robust OFDMA scheme. } \vspace{-0.3cm}
    \label{roma}
    \vspace{0.5cm}
    \includegraphics[scale=0.28]{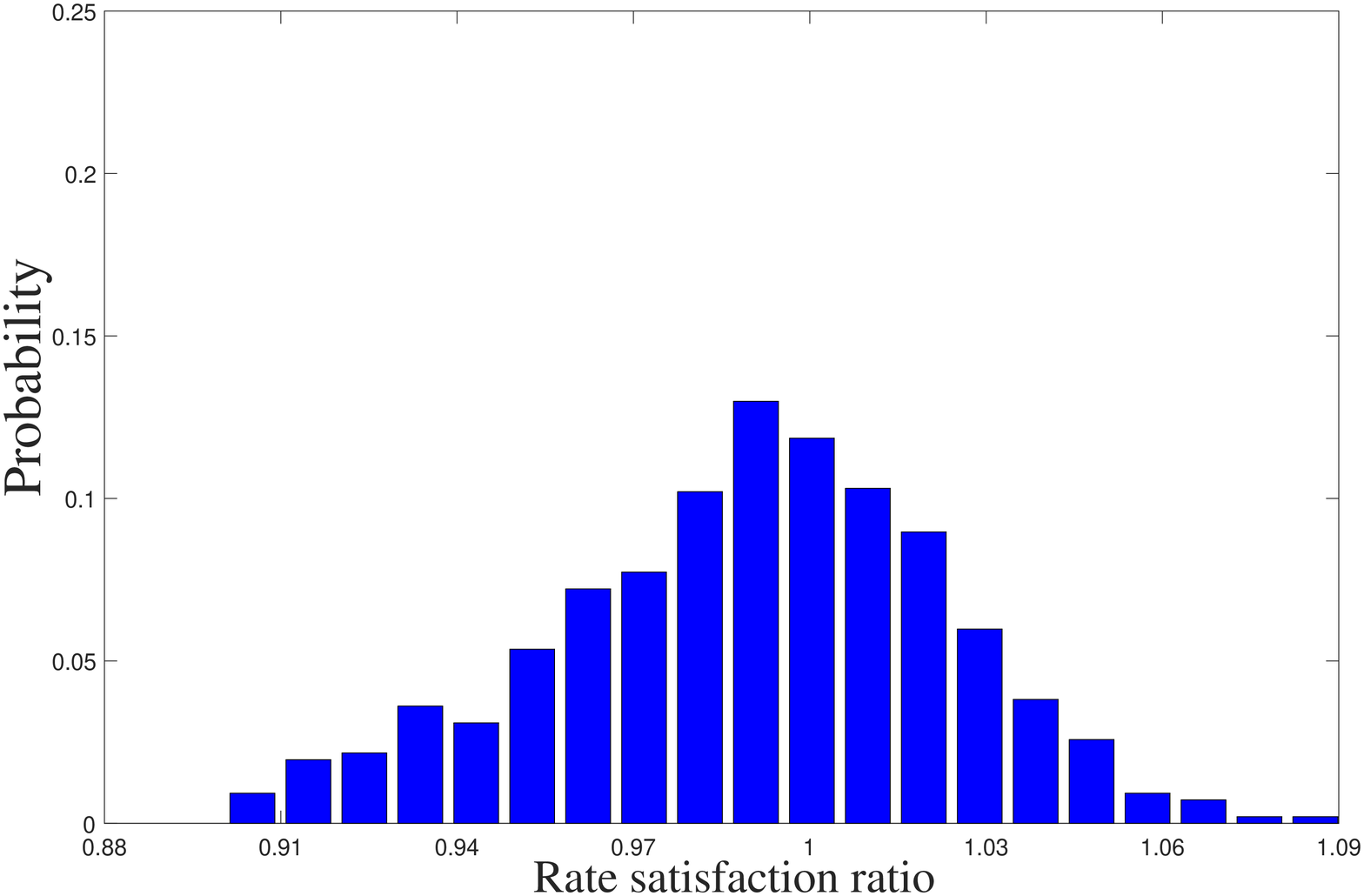} 
    \vspace{-0.6cm}
    \caption{Histogram for rate satisfaction ratio, i.e., $\eta_k$,  for $R_{min}=3 bps/Hz$ in the non-robust OFDMA scheme. }\vspace{-0.3cm}
    \label{nroma}
  \end{center}
\end{figure}

\begin{table*}[th!]
\caption{Comparison of power allocations between Taylor series approximation and SDR approaches}\label{table1}
\centering
\resizebox{\textwidth}{!}{
\begin{tabular}{|c|c|c|c|c|c|c|c|c|}
\cline{2-9} \multicolumn{1}{c|}{} & \multicolumn{4}{c|}{Taylor Series Approximation Scheme} & \multicolumn{4}{c|}{SDR Scheme} \\
\hline
Channels
&$\begin{array}{c}
             U_1 ~\text{Power} \\
            ({\rm w})
           \end{array}$& $\begin{array}{c}
             U_2 ~\text{Power} \\
            ({\rm w})
           \end{array}$& $\begin{array}{c}
             U_3 ~\text{Power} \\
            ({\rm w})
           \end{array}$& $\begin{array}{c}
             \text{Total Power}\\
            ({\rm w})
           \end{array}$& $\begin{array}{c}
             U_1 ~\text{Power} \\
            ({\rm w})
           \end{array}$& $\begin{array}{c}
             U_2 ~\text{Power} \\
            ({\rm w})
           \end{array}$& $\begin{array}{c}
             U_3 ~\text{Power} \\
            ({\rm w})
           \end{array}$&$\begin{array}{c}
             \text{Total Power}\\
            ({\rm w})
           \end{array}$ \\
\hline
Channel $1$  &5.3596&	0.6702&	0.0821&	6.1119&	5.3593&	0.6701&	0.0820&	6.1114\\
\hline
Channel $2$  &3.7419&	0.4692&	0.0636	&4.2747&	3.7417&	0.4690&	0.0635&	4.2742\\
\hline
Channel $3$  &7.6185&	0.9595&	0.1232&	8.7012&	7.6156&	0.9591&	0.1230&	8.6977\\
\hline
Channel $4$  &10.2415&	1.2811&	0.1642&	11.6868&	10.2367&	1.2808&	0.1641&	11.6814\\
\hline
Channel $5$  &8.4636&	1.0626&	0.1468&	9.6730&	8.4634&	1.0625&	0.1467&	9.6726\\
\hline
\end{tabular}}
\end{table*}

\begin{table*}[th!]
\caption{Power allocations and achieved rate of the max-min fairness approach}\label{table2}
\centering
\resizebox{\textwidth}{!}{
\begin{tabular}{|p{1.6cm} | c| c| c|c| c| c| c|}
\hline
Channels & $\begin{array}{c}
             U_1 ~\text{Power} \\
            ({\rm w})
           \end{array}$
     & $\begin{array}{c}
             U_2 ~\text{Power} \\
             ({\rm w})
           \end{array}$ & $\begin{array}{c}
             U_3 ~\text{Power} \\
             ({\rm w})
           \end{array}$  & $\begin{array}{c}
             \text{Total Power}\\
             ({\rm w})
           \end{array}$& $\begin{array}{c}
             U_1 ~\text{Rate} \\
             (bps/Hz)
           \end{array}$& $\begin{array}{c}
             U_2 ~\text{Rate} \\
             (bps/Hz)
           \end{array}$ & $\begin{array}{c}
             U_3 ~\text{Rate} \\
             (bps/Hz)
           \end{array}$ \\
\hline
Channel $1$  & 13.9225& 0.9990&  0.0804&	15&	3.8010&	3.8010&	3.8010\\
\hline
Channel $2$ & $4.3137$ & $0.5965$ & $0.0929$ & $5$ & $2.8584$ & $2.8584$ & $2.8584$ \\
\hline
Channel $3$ & 9.1271& 0.8013&	0.0730&	10&	3.5140&	3.5140&	3.5140\\
\hline
Channel $4$ & 6.7110&	0.7166&	0.0754&	7.5&	3.2327&	3.2327&	3.2327\\
\hline
Channel $5$ & 11.5454& 0.8814& 0.0752&	12.5&	3.7119&	3.7119&	3.7119\\
\hline
\end{tabular}
}
\end{table*}

\begin{table*}[th!]
\caption{{Power allocations for a given set target rates (i.e., achieved in fairness approach) through power minimization in} \eqref{mainproblem2} }\label{table3}
\centering
\resizebox{\textwidth}{!}{
\begin{tabular}{|p{2cm} | c| c| c| c| c|}
\hline
Channels  & $\begin{array}{c}
             \text{Target Rate} \\
            ({\rm bps/Hz})
           \end{array}$
           & $\begin{array}{c}
             \text{Total Power} \\
             ({\rm w})
           \end{array}$
           & $\begin{array}{c}
             U_1 ~\text{Power} \\
             ({\rm w})
           \end{array}$
           & $\begin{array}{c}
             U_2 ~\text{Power} \\
             ({\rm w})
           \end{array}$
            & $\begin{array}{c}
             U_3 ~\text{Power} \\
             ({\rm w})
           \end{array}$\\
\hline
Channel $1$  &3.8010 &15& 13.9225& 0.9990&  0.0804	\\
\hline
Channel $2$  & 2.8584& 5& 4.3137 & 0.5965 & 0.0929 \\
\hline
Channel $3$  & 3.5140 &10& 9.1271& 0.8013&	0.0730 \\
\hline
Channel $4$ &	3.2327&	7.5& 6.7110&	0.7166&	0.0754 \\
\hline
Channel $5$  &	3.7119 &	12.5& 11.5454& 0.8814& 0.0752 \\
\hline
\end{tabular}
}
\end{table*}

\appendices \numberwithin{equation}{section}
\setcounter{equation}{0}

\section{Proof of Lemma 1} \label{App.A}
{We first summarize the following complex derivatives:}

Generalized complex derivative:

\begin{equation}\label{1}
 \frac{\partial f(z)}{\partial z}=\frac{1}{2}\big( \frac{\partial f(z)}{\partial \Re(z)}-i\frac{\partial f(z)}{\partial \Im(z)}\big)
\end{equation}

Conjugate complex derivative:

\begin{equation}\label{2}
 \frac{\partial f(z)}{\partial z^*}=\frac{1}{2}\big( \frac{\partial f(z)}{\partial \Re(z)}+i\frac{\partial f(z)}{\partial \Im(z)}\big)
\end{equation}

\begin{equation}\label{3}
  \Rightarrow \qquad \frac{\partial f(z)}{\partial \Re(z)}=2 \Re[ \frac{\partial f(z)}{\partial z^*}],\qquad \frac{\partial f(z)}{\partial \Im(z)}=2 \Im[ \frac{\partial f(z)}{\partial z^*}]
\end{equation}

{Next, we present the first order Taylor series approximation for a function $g(A)$ around $A_0$ as follows:}

\begin{align}\nonumber
  g(A) \approx g(A,A_0) =& f(A_0)+ {\rm Tr} \Big( \big[\frac{\partial f(A_0)}{\partial \Re(A)}\big]^T \Re(A-A_0)\Big) \\ \nonumber
  &+{\rm Tr} \Big( \big[\frac{\partial f(A_0)}{\partial \Im(A)}\big]^T \Im(A-A_0)\Big)\\\nonumber
   =& f(A_0)+ {\rm Tr} \Big( 2 \big[\frac{\partial f(A_0)}{\partial A^*}\big]^T \Re(A-A_0)\Big) \\ \nonumber
   &+{\rm Tr} \Big( 2 \big[\frac{\partial f(A_0)}{\partial A^* }\big]^T \Im(A-A_0)\Big) \\
   =&  f(A_0)+ 2 \Re\Big( {\rm Tr} \big[\frac{\partial f(A_0)}{\partial A^*}\big]^H (A-A_0)\Big)
\end{align}

Based on this first order approximation, we approximate
$ f_l(\mathbf{w}_{k}) = \mathbf{w}^H_k \mathbf{h}_l \mathbf{h}_l^H \mathbf{w}_k$ as follows:
\begin{align} \nonumber
  g_l(\mathbf{w}_{k},\mathbf{w}^t_{k} ) =& {\mathbf{w}^t_{k}}^H \mathbf{h}_l{\mathbf{h}_l}^H \mathbf{w}^t_k \\ \nonumber
  &+\bigg[ \Big( \frac{\partial f_l(\mathbf{w}_{k}) }{\partial \Re(\mathbf{w}_{k})}\Big|_{\mathbf{w}_{k}=\mathbf{w}^t_{k}}\Big)^T
\Re(\mathbf{w}_{k}-\mathbf{w}^t_{k}) \bigg]\\ \label{5}
&+
\bigg[ \Big( \frac{\partial f_l(\mathbf{w}_{k}) }{\partial \Im(\mathbf{w}_{k})}\Big|_{\mathbf{w}_{k}=\mathbf{w}^t_{k}}\Big)^T
\Im(\mathbf{w}_{k}-\mathbf{w}^t_{k}) \bigg]
\end{align}

{The derivatives of the following terms can be written based on the complex derivatives in \eqref{3}:}
\begin{equation}\label{6}
  \frac{\partial f_l(\mathbf{w}_{k}) }{\partial \Re(\mathbf{w}_{k})} = 2 \Re(\mathbf{h}_l{\mathbf{h}_l}^H \mathbf{w}_{k}), ~
   \frac{\partial f_l(\mathbf{w}_{k}) }{\partial \Im(\mathbf{w}_{k})} = 2 \Im(\mathbf{h}_l{\mathbf{h}_l}^H \mathbf{w}_{k}),
\end{equation}

\begin{eqnarray} \nonumber
   g_l(\mathbf{w}_{k},\mathbf{w}^t_{k} ) &=&
   {\mathbf{w}^t_{k}}^H \mathbf{h}_l \mathbf{h}_l^H \mathbf{w}^t_{k} \\\nonumber
&&+  2 \big(\Re( {\mathbf{w}^t_{k}}^H\mathbf{h}_l{\mathbf{h}_l}^H)\Re(\mathbf{w}_{k}-\mathbf{w}^t_{k}) \big)\\ \nonumber
&& +  2 \big(\Im( {\mathbf{w}^t_{k}}^H\mathbf{h}_l{\mathbf{h}_l}^H)\Im(\mathbf{w}_{k}-\mathbf{w}^t_{k}) \big)\\\nonumber
&=& {\mathbf{w}^t_{k}}^H \mathbf{h}_l \mathbf{h}_l^H \mathbf{w}^t_{k} \\
&&+ 2 \Re[{\mathbf{w}^t_k}^H \mathbf{h}_l{\mathbf{h}_l}^H(\mathbf{w}_{k}-\mathbf{w}^t_{k})]
\end{eqnarray}

This completes the proof of Lemma 1. \qquad\qquad\qquad\,\: $\blacksquare$

\section{Proof of Lemma 3} \label{App.C}
In order to convert the probability based constraints in \eqref{robustproblem2} into a
tractable form the following lemma is required:

\textit{Lemma 3.1:} Consider a hermitian random
matrix $\mathbf{X}\in \mathbb{C}^{M\times M}$ with each element being independent and identically distributed as $[\mathbf{X}]_{ij}\sim
\mathcal{CN}(0, \sigma^2_{ij})$. Then, for any hermitian matrix \mbox{$\mathbf{Y}\in \mathbb{C}^{M\times M}$ }, the following holds:
\begin{align} \nonumber
  &\rm{Tr}(\mathbf{YX})\sim \mathcal{CN}(0, \|\mathbf{Y} \odot \Sigma_X\|^2_F),\\ \nonumber
  &\rm{Tr}(\mathbf{YX})=\|\mathbf{Y} \odot \Sigma_X\|_F U,~ U \sim \mathcal{N}(0, 1),
\end{align}
where $\odot$ indicates the Hadamard product and $\Sigma_X$  represents a real valued $M\times M $ matrix with each entry $[\Sigma_X]_{ij}=\sigma_{ij}$.  

By defining a new rank-one positive semidefinite matrix
$\mathbf{W}_k=\mathbf{w}_k \mathbf{w}_k^H$, the constraints in \eqref{robustproblem2} can be rewritten, respectively, as follows:
\begin{align}\nonumber
&{\rm Pr}\Big( {\rm Tr}(-\mathbf{B}_{k} \mathbf{\Delta}_l)\leq {\rm Tr}(\mathbf{B}_{k} \mathbf{\hat{C}}_l)-\sigma^2\Big)\geq (1-\rho_k),\\ \label{simplifiedRobust}
&\hspace{1.5 cm}\forall k, l=k,k+1,\ldots,N.
\end{align}
where $\mathbf{B}_k=\gamma_k^{-1}\mathbf{W}_k-\sum_{m=k+1}^N
\mathbf{W}_m$.

By exploiting Lemma 3.1 and the cumulative distribution function (CDF) of a standard normal distribution,
(i.e., ${\rm Pr}(U \leq u) = \frac{1}{2}[1+ {\rm erf}(\frac{u}{\sqrt{2}})]$, where $\mathbf{U}\sim \mathcal{N}(0,1)$ ), the inequalities
\eqref{simplifiedRobust} can be represented as follows:

\begin{align}\nonumber
&{\rm Pr}\Big( {\rm Tr}(-\mathbf{B}_k \mathbf{\Delta}_{l})\leq {\rm Tr}( \mathbf{B}_k \mathbf{\hat{C}}_{l})-\sigma^2\Big) \\ \nonumber
&= {\rm Pr} \Big( U \leq \frac{{\rm Tr}( \mathbf{B}_k \mathbf{\hat{C}}_{l})-\sigma^2}{\|-\mathbf{B}_k \odot \Sigma_{\mathbf{\Delta}_{l}}\|_F}\Big) \\ \nonumber
&= \frac{1}{2}[1+{\rm erf}(\frac{{\rm Tr}( \mathbf{B}_k \mathbf{\hat{C}}_{l})-\sigma^2}{\sqrt{2} \|-\mathbf{B}_k \odot \Sigma_{\mathbf{\Delta}_{l}}\|_F})]  \\ \nonumber
&\geq (1-\rho_k), \\ \label{firstPr}
& \forall k,~l=k,k+1,\ldots,N,
\end{align}

The inequalities in \eqref{firstPr} can be presented in the following forms,
\begin{align}\label{lemm1.1}
\Phi_{kl} \geq \sqrt{2}~ {\rm erf}^{-1}(1-2\rho_k)\| {\rm vec}(-\mathbf{B}_k \odot \Sigma_{\mathbf{\Delta}_{l}}) \|,
\end{align}
 where
\begin{align}
& \Phi_{kl}={\rm Tr}( \mathbf{B}_k \mathbf{\hat{C}}_{l})-\sigma^2.
\end{align}

In order to cast the original robust  problem into a convex optimization framework, the following lemma is required:

\textit{Lemma 3.2:} The following second order cone \mbox{constraint on $x$}
\begin{equation} \nonumber
  \|\mathbf{A}x+b\|\leq e^T x + d,
\end{equation}
can be represented with the following linear matrix inequality (LMI): 
\begin{equation}\nonumber
  \left [
    \begin{array}{cc}
      (e^T x + d)\mathbf{I} & \mathbf{A}x+b \\
      (\mathbf{A}x+b)^T & e^T x + d \\
    \end{array}
  \right ]\succeq 0 .
\end{equation}

By applying Lemma 3.2, the constraints \eqref{lemm1.1} can be rewritten as
\begin{align}\nonumber
\mathbf{C}_{kl}&=\left [
    \begin{array}{cc}
      \frac{\Phi_{kl}}{\sqrt{2} {\rm erf}^{-1}(1-2\rho_k)}\mathbf{I}_{M^2} & {\rm vec}(-\mathbf{B}_k \odot \Sigma_{\mathbf{\Delta}_{l}}) \\
      {\rm vec}^H(-\mathbf{B}_k \odot \Sigma_{\mathbf{\Delta}_{l}}) & \frac{\Phi_{kl}}{\sqrt{2} {\rm erf}^{-1}(1-2\rho_k)} \\
    \end{array}
  \right ], \\\label{con1}
  &\qquad\qquad \forall k,~l=k,k+1,\ldots,N.
\end{align}

\noindent This completes the proof of Lemma 3. \qquad\qquad\qquad\quad $\blacksquare$

\vspace{-0.3cm}
\bibliographystyle{IEEEtrans}
\bibliography{IEEEabrv,Bibliography}
\end{document}